\documentclass
[preprint,pra,showpacs,showkeys,byrevtex,nobibnotes,nofootinbib,12pt,a4paper,onecolumn]{revtex4}%
\usepackage{amsfonts}
\usepackage{amsmath}
\usepackage{amssymb}
\usepackage{graphics}
\usepackage{graphicx}%
\providecommand{\U}[1]{\protect\rule{.1in}{.1in}}

\begin{document}
\title{On a possible definition of the moving preferred basis}
\author{Mario Castagnino}
\affiliation{CONICET, IAFE (CONICET-UBA), IFIR and FCEN (UBA), Argentina.}
\author{Sebastian Fortin}
\affiliation{CONICET, IAFE (CONICET-UBA) and FCEN (UBA), Argentina.}

\begin{abstract}
There are many formalisms to describe quantum decoherence. However, many of
them give a non general and \ ad hoc definition of \textquotedblleft pointer
basis\textquotedblright\ or \textquotedblleft moving preferred
basis\textquotedblright, and this fact is a problem for the decoherence
program. In this paper we will consider quantum systems under a general
theoretical framework for decoherence and present a tentative very general
definition of the moving preferred basis. which is implemented in a well known
open system model. The obtained decoherence and the relaxation times are
defined and compared with those of this model.

\end{abstract}
\maketitle
\tableofcontents

\section{Introduction}

From the appearance of the quantum mechanics many attempts have been made to
recover the laws of the classic mechanics through some classic limit. The more
common scheme of this type includes the \textit{quantum decoherence}%
\footnote{We will call \textit{decoherence} to the vanishing of the
off-diagonal terms in a properly specified basis. We will call
\textit{relaxation} to the decoherence in a final equilibrium basis, i.e.
typical equilibrium}. This process is in charge to erase the terms of
interference of the density matrix, that are classically inadmissible, since
they prevent the use of a classical (boolean) logic. In addition, decoherence
leads to the rule that selects the candidates for classic states.

As it is pointed out in the brief historical summary of paper
\cite{CQG-CFLL-08}, three periods can be schematically identified in the
development of the general program of decoherence \cite{Omnes}. A first
period, when the arrival to the equilibrium of irreversible systems was
studied. During this period, authors as van Kampen, van Hove, Daneri, et
al\textit{.} developed a formalism for explaining the decoherence phenomenon
that was not successful at the time but it established the bases of this
study. The main problem of this period was that too long \textit{decoherence
times }$t_{D}$ were found, if compared with the experimental ones (which was
of the order of the time in which the systems reach equilibrium, i. e.
\textit{relaxation time }$t_{R}).$ In a second period the decoherence in open
systems was studied, the main characters of this period were Zeh and Zurek. In
their works, the decoherence is an interaction process between an open quantum
system and its environment. This process, called \textit{Environment-Induced
Decoherence} (EID), determines, case by case, which is the privileged basis,
usually called \textit{moving preferred basis} where decoherence takes place
in a decoherence time $t_{D}\ll$ $t_{R}$ and it defines the observables that
acquire classic characteristics and they could be interpreted in some
particular cases as properties that obey a Boolean logic. This is the orthodox
position in the subject \cite{Bub}. So decoherence times in this period were
much smaller, solving the problem of the first period. Recently, in a third
period it becomes evident that dissipation was not a necessary condition for
decoherence \cite{Max} and the study of the arrival to equilibrium of closed
systems was also considered. We will not discuss closed systems in this paper
but for the sake of completeness we will make only some comments. Closed
system will be discussed at large elsewhere.

In this work we focus the attention on EID, which is a well known theory, with
well established experimental verifications, which makes unnecessary any
further explanation. On the contrary other formalisms are not so well
established, but they must be taken into account for the sake of completeness
(\cite{GP}, \cite{connel}, \cite{CP}, \cite{SID}, \cite{SID'}, \cite{DT},
\cite{MHI}, \cite{PLA}, \cite{JPA'}, \cite{Studies}).

In this paper, we will introduce a tentative definition of the moving
preferred basis where the state decoheres in a very short time $t_{D}$, So the
main problem of the first period is solved in a convenient and general way.
Our main aim is to present a new conceptual perspective that will clarify some
points that still remain rather obscure in the literature on the subject, (e.
g. the definition of the moving preferred basis and the use of the pole technique)

\subsection{The General Theoretical Framework for Decoherence}

In previous works we have resumed the common characteristics of the different
approaches of decoherence, which suggest the existence of a general framework
for decoherence within which these approaches can all be framed (see
\cite{CQG-CFLL-08}, \cite{Studies} and \cite{MPLA}). According to this general
framework, that was developed in \cite{Studies}, and will be completed in
future papers, decoherence is just a particular case of the general problem of
irreversibility in quantum mechanics. Since the quantum state $\rho(t)$
follows a unitary evolution, it cannot reach a final equilibrium state for
$t\rightarrow\infty$. Therefore, if the non-unitary evolution towards
equilibrium is to be accounted for, a further element has to be added to this
unitary evolution. The way to introduce this non-unitary evolution must
include the splitting of the whole space of observables $\mathcal{O}$ into the
relevant subspace $\mathcal{O}_{R}\subset\mathcal{O}$ and the irrelevant
subspace. Once the essential role played by the selection of the relevant
observables is clearly understood, the phenomenon of decoherence can be
explained in four general steps:

\begin{enumerate}
\item \textbf{First step:} The space $\mathcal{O}_{R}$ of relevant observables
is defined .

\item \textbf{Second step:} The expectation value $\langle O_{R}\rangle
_{\rho(t)}$, for any $O_{R}\in\mathcal{O}_{R}$, is obtained. This step can be
formulated in two different but equivalent ways:

\begin{itemize}
\item A coarse-grained state $\rho_{R}(t)$ is defined by
\begin{equation}
\langle O_{R}\rangle_{\rho(t)}=\langle O_{R}\rangle_{\rho_{R}(t)} \label{0}%
\end{equation}
for any $O_{R}\in\mathcal{O}$, and its non-unitary evolution (governed by a
master equation) is computed (this step is typical in EID).

\item $\langle O_{R}\rangle_{\rho(t)}$ is computed and studied as the
expectation value of $O_{R}$ in the state $\rho(t)$. This is the generic case
for other formalisms.
\end{itemize}

\item \textbf{Third step:} It is proved that $\langle O_{R}\rangle_{\rho
(t)}=\langle O_{R}\rangle_{\rho_{R}(t)}$ reaches a final equilibrium value
$\langle O_{R}\rangle_{\rho_{\ast}}$, then
\begin{equation}
\lim_{t\rightarrow\infty}\langle O_{R}\rangle_{\rho(t)}=\langle O_{R}%
\rangle_{\rho_{\ast}},\text{ \ \ \ \ \ \ \ \ \ \ \ }\forall O_{R}%
\in\mathcal{O}_{R} \label{INT-01}%
\end{equation}

This also means that the coarse-grained state $\rho_{R}(t)$ evolves towards a
final equilibrium state:%
\begin{equation}
\lim_{t\rightarrow\infty}\langle O_{R}\rangle_{\rho_{R}(t)}=\langle
O_{R}\rangle_{\rho_{R\ast}},\ \ \ \ \ \ \ \ \ \ \ \forall O_{R}\in
\mathcal{O}_{R} \label{INT-02'}%
\end{equation}

The characteristic time for these limits is the $t_{R}$, the
\textit{relaxation time.}

\item \textbf{Fourth step: }Also a \textit{moving preferred basis}
$\{|\widetilde{j(t)\rangle}\}$ must be defined as we will see in section I.B.
This basis is the eigen basis of certain state $\rho_{P}(t)$ such that%
\begin{equation}
\lim_{t\rightarrow\infty}\langle O_{R}\rangle_{(\rho_{R}(t)-\rho_{P}%
(t))}=0,\ \ \ \ \ \ \ \ \ \ \ \forall O_{R}\in\mathcal{O}_{R}%
\end{equation}
The characteristic time for this limit is the $t_{D}$, the \textit{decoherence
time.}
\end{enumerate}

The final equilibrium state $\rho_{\ast}$ is obviously diagonal in its own
eigenbasis, which turns out to be the final preferred basis. But, from eqs.
(\ref{INT-01}) or (\ref{INT-02'}) we cannot say that $\lim_{t\rightarrow
\infty}\rho(t)=\rho_{\ast}$ or $\lim_{t\rightarrow\infty}\rho_{R}%
(t)=\rho_{R\ast}.$ Then, the mathematicians say that the unitarily evolving
quantum state $\rho(t)$ of the whole system \textit{only has a} \textit{weak
limit, }symbolized as:
\begin{equation}
W-\lim_{t\rightarrow\infty}\rho(t)=\rho_{\ast} \label{INT-03}%
\end{equation}
equivalent to eq. (\ref{INT-01}). As a consequence, the coarse-grained state
$\rho_{R}(t)$ also has a weak limit, as follows from eq.(\ref{INT-02'}):
\begin{equation}
W-\lim_{t\rightarrow\infty}\rho_{R}(t)=\rho_{R\ast} \label{INT-04}%
\end{equation}
equivalent to eq. (\ref{INT-02'}). Also%
\begin{equation}
W-\lim_{t\rightarrow\infty}(\rho_{R}(t)-\rho_{P}(t))=0
\end{equation}
These weak limits mean that, although the off-diagonal terms of $\rho(t)$
never vanish through the unitary evolution, the system decoheres \textit{from
an observational point of view}, that is, from the viewpoint given by any
relevant observable $O_{R}\in\mathcal{O}_{R}$.

From this general perspective, the phenomenon of destructive interference,
that produced the decoherence phenomenon is relative, because the off-diagonal
terms of $\rho(t)$ and $\rho_{R}(t)$ vanish only from the viewpoint of the
relevant observables\textbf{ }$O_{R}\in\mathcal{O}_{R}$, and the
superselection rule that precludes superpositions only retains the states
defined by the corresponding decoherence bases as we will see. The only
difference between EID and other formalisms for decoherence is the selection
of the relevant observables (see \cite{CQG-CFLL-08} for details):

\begin{description}
\item[.] In EID the relevant observables are those having the following form:%
\begin{equation}
O_{R}=O_{S}\otimes I_{E}\in\mathcal{O}_{R} \label{EID}%
\end{equation}
where $O_{S}$ are the observables of the system and $I_{E}$ is the identity
operator of the environment. Then eq. (\ref{0}) reads%
\[
\langle O_{R}\rangle_{\rho(t)}=\langle O_{R}\rangle_{\rho_{R}(t)}=\langle
O_{S}\rangle_{\rho_{S}(t)},\text{ where }\rho_{S}(t)=Tr_{E}\rho(t)
\]
where in $Tr_{E}\rho(t)$ we have "traced away" the environment. In the other
formalisms other restriction in the set of observables can be introduced
\end{description}

\subsection{The definition of moving preferred basis}

The moving preferred basis was introduced, case by case in several papers (see
\cite{Max}) in a non systematic way. On the other hand in references
\cite{OmnesPh} and \cite{OmnesRojo} Roland Omn\`{e}s introduces a rigorous and
almost general definition of the moving preferred basis based in a reasonable
choice of the relevant observables, and other physical considerations.

In this paper we will introduce an alternative general definition to define
this basis: As it is well known the eigen values of the Hamiltonian are the
inverse of the characteristic frequencies of the unitary evolution of an
oscillatory system. Analogously, for non-unitary evolutions, the poles of the
complex extension of the Hamiltonian are the \textit{catalogue} of the
decaying modes of these non-unitary evolutions towards equilibrium (see
\cite{JPA}). This will be the main idea to implement the definition of our
moving preferred basis. i. e. we will use only the poles nearest to the real
axis and we will eliminate the other poles in order to obtain an
adiabatic-like definition.

We will compare and try to unify these two methods in the future. Really we
already have begin this approach with Omn\`{e}s in section III.

\subsection{Organization of the paper.}

. In Section I we have introduced a general framework for decoherence. Some
candidates for moving preferred basis are introduced in section
\ref{GenDefMovDecBas}, which is implemented in two models and the times of
decoherence and relaxation and the moving preferred basis in these models are
defined. In principle these definitions can be used in EID and probably for
other formalisms. In Section III we will define the moving preferred basis of
our formalism and we will present the paradigmatic EID: Omn\`{e}s (or
Lee-Friedrich) model. and show that the pole method yields the usual results.
Finally in Section \ref{Conclusions} we will draw our conclusions. An appendix
completes this paper.

\section{\label{GenDefMovDecBas}Towards a general definition for the moving
preferred basis.}

\subsection{Introduction and review}

In this section we will try to introduce a very general theory for the moving
preferred basis in the case of a general distribution of poles and for
\textit{any relevant observable space} $\mathcal{O}_{R}.$ Then it is necessary
to endow the coordinates of observables and states in the Hamiltonian basis
$\{|\omega\rangle\}$ (i.e. the functions $O(\omega,\omega^{\prime})$ and
$\rho(\omega.\omega^{\prime}))$ with extra analytical properties in order to
find the definition of a moving preferred basis in the most, general,
convincing, and simplest way. It is well known that this move is usual in many
chapters of physic e. g. in the scattering theory (see \cite{Bohm}).

It is also well known that evolution towards equilibrium has two phases.

i.- A exponential dumping phase that can be described studying the analytical
continuation of the Hamiltonian into the complex plane of the energy (see
\cite{JPA}, \cite{Weder}, \cite{Sudarsham}, \cite{PRA1}, \cite{PRA2},
\cite{PRE}), a fact which is also well known in the scattering theory.

ii.- A final decaying inverse-polynomial in $t^{-1}$ known as the long time of
Khalfin effect (see \cite{Khalfin}, \cite{BH}), which is very weak and
difficult to detect experimentally (see \cite{KhalfinEx})\footnote{There is
also an initial ($t=0)$ non exponential Zeno-period which is unimportant for
this paper}.

These two phases will play an important role in the definition of the moving
preferred basis. They can be identified by the theory of analytical
continuation of vectors, observables and states. To introduce the main
equations we will make a short abstract of papers \cite{JPA} and \cite{PRA2}.

\subsection{Analytic continuations in the bra-ket language.}

We begin reviewing the analytical continuation for pure states. Let the
Hamiltonian be $H=H_{0}+V$ where the free Hamiltonian $H_{0}$ satisfies (
\cite{JPA}. eq. (8) or \cite{PRA2})%

\[
H_{0}|\omega\rangle=\omega|\omega\rangle,\text{ }\langle\omega|H_{0}%
=\omega\langle\omega|,\text{ \ \ }0\leq\omega<\infty
\]
and (see \cite{JPA}. eq. (9))%
\begin{equation}
I=\int_{0}^{\infty}d\omega|\omega\rangle\langle\omega|,\text{ }\langle
\omega|\omega^{\prime}\rangle=\delta(\omega-\omega^{\prime}) \label{I}%
\end{equation}
Then (see \cite{JPA}. eq. (10))
\[
H_{0}=\int_{0}^{\infty}\omega|\omega\rangle\langle\omega|d\omega
\]
and (see \cite{JPA}. eq. (11))
\begin{equation}
H=H_{0}+V=\int_{0}^{\infty}\omega|\omega\rangle\langle\omega|d\omega+\int
_{0}^{\infty}d\omega\int_{0}^{\infty}d\omega^{\prime}V_{\omega\omega^{\prime}%
}|\omega\rangle\langle\omega^{\prime}|=\int_{0}^{\infty}\omega|\omega
^{+}\rangle\langle\omega^{+}|d\omega\label{Hamil}%
\end{equation}
where the $|\omega^{+}\rangle$ are the eigenvectors of $H,$ that also satisfy
eq. (\ref{I}). The eigen vectors of $H$ are given by the Lippmann-Schwinger
equations (see \cite{JPA}. eq. (12) and (13))%
\begin{equation}
\langle\psi|\omega^{+}\rangle=\langle\psi|\omega\rangle+\langle\psi|\frac
{1}{\omega+i0-H}V|\omega\rangle,\text{ \ \ }\langle\omega^{+}|\varphi
\rangle=\langle\omega|\varphi\rangle+\langle\omega|V\frac{1}{\omega
-i0-H}|\varphi\rangle\label{AN}%
\end{equation}
Let us now endow the function of $\omega$ with adequate analytical properties
(see \cite{Bohm}). E.g. let us consider that the state $|\varphi\rangle$
(resp. $\langle\psi|)$ is such that it does not create poles in the complex
extension of $\langle\omega|\varphi\rangle$ (resp. in $\langle\psi
|\omega\rangle)$ and therefore this function is analytic in the whole complex
plane. This is a simplification that we will be forced to abandon in some
cases as we will see in remark 1 of section II.E. Moreover we will consider
that the complex extensions of function $\langle\omega^{+}|\varphi\rangle$
(resp. $\langle\psi|\omega^{+}\rangle)$ is analytic but with just one simple
pole at $z_{0}$ $=\omega_{0}-\frac{i}{2}\gamma_{0},$ $\gamma_{0}>0$ in the
lower halfplane (resp. another pole $z_{0}^{\ast}=\omega_{0}+\frac{i}{2}%
\gamma_{0},\gamma_{0}>0$ on the upper halfplane$)$ (see \cite{DT} for details
\footnote{This is a toy model with just one pole and the Khalfin effect. More
general models , with two poles, will be considered in the next subsection.
The pole corresponds to the residue that we can compute with the curve
\textit{C} and the Khalfin effect to the integral along the curve $\Gamma$ of
Figure 1.}). There can be many of such poles but , by now, we will just
consider one pole for simplicity, being the generalization straightforward.
Then we make an analytic continuation of the positive $\omega$ axis to the
curve $\Gamma$ of the lower half-plane as in Figure 1\footnote{All the figures
are merely illustrative and not the numerical solution of some precise
problem.}.%

\begin{figure}[t]
\centerline{\scalebox{0.7}{\includegraphics{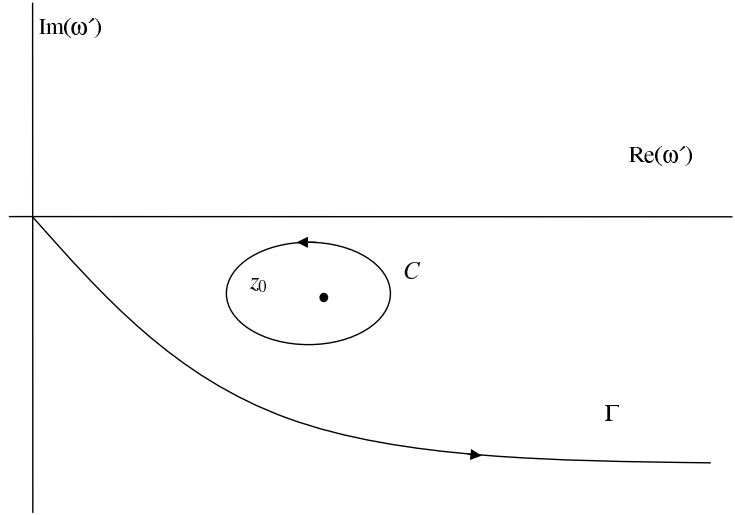}}} \vspace
*{0.cm}\caption{Complex contour $\Gamma$ on the lower complex energy plane
usedin our evaluation of integrals. The \textquotedblleft
energy\textquotedblright\ $z_{0}$ is the pole that we assume to be simple.}%
\label{fig 1}%
\end{figure}

Then (see \cite{JPA}. eq. (29)) we can define%
\[
\langle\widetilde{f_{0}}|\varphi\rangle\equiv cont_{\omega^{\prime}\rightarrow
z_{0}}\langle\omega^{\prime+}|\varphi\rangle,\text{ \ \ }\langle\psi
|f_{0}\rangle\equiv(-2\pi i)cont_{\omega^{\prime}\rightarrow z_{0}}%
(\omega^{\prime}-z_{0})\langle\psi|\omega^{+}\rangle
\]%
\begin{equation}
\langle\widetilde{f_{z^{\prime}}}|\varphi\rangle\equiv cont_{\omega^{\prime
}\rightarrow z^{\prime}}\langle\omega^{\prime+}|\varphi\rangle,\text{
\ \ }\langle\psi|f_{z^{\prime}}\rangle\equiv cont_{\omega^{\prime}\rightarrow
z}\langle\psi|\omega^{+}\rangle,\text{\ }z^{\prime}\in\Gamma,\forall\text{
}|\varphi\rangle\text{ }\langle\psi| \label{L.1}%
\end{equation}
and (see \cite{JPA}. eq. (31))%
\[
\langle\psi|\widetilde{f_{0}}\rangle\equiv cont_{\omega\rightarrow z_{0}%
^{\ast}}\langle\psi|\omega^{+}\rangle,\text{ \ \ }\langle f_{0}|\varphi
\rangle\equiv(2\pi i)cont_{\omega^{\prime}\rightarrow z_{0}^{\ast}}%
(\omega-z_{0})\langle\omega^{+}|\varphi\rangle
\]%
\begin{equation}
\langle\psi|\widetilde{f_{z^{\prime}}}\rangle\equiv cont_{\omega\rightarrow
z}\langle\psi|\omega^{+}\rangle,\text{ \ \ }\langle f_{z}|\varphi\rangle\equiv
cont_{\omega\rightarrow z}\langle\omega^{+}|\varphi\rangle,\text{ \ }%
z\in\Gamma,\forall\text{ }|\varphi\rangle\text{ }\langle\psi| \label{L.2}%
\end{equation}
where $cont$ means analytic continuation.

Finally it can be proved that (see \cite{JPA})
\[
H=z_{0}|f_{0}\rangle\langle\widetilde{f_{0}}|+\int_{\Gamma}z|f_{z}%
\rangle\langle\widetilde{f_{z}}|dz
\]
a simple extension of the eigen-decomposition of $H$ to the complex plane

\subsection{Analytical continuation in the observables and states language.}

We could repeat what we have said about the pure states and the Hamiltonian
with the states, observables, and the Liouvillian operator $\ L$ (see a review
in \cite{Cos}). But we prefer to follow the path of \cite{JPA} and keep the
Hamiltonian framework and discuss the analytical continuation of \ $\langle
O\rangle_{\rho(t)},$ that we will also symbolize as $(\rho(t)|O)$. In fact
from section I.A we know that this scalar is the main character of the play so
we will study its analytical properties ad nauseam.

So let us call (see \cite{JPA}. eq. (42))
\[
|\omega)=|\omega\rangle\langle\omega|,\text{ and }|\omega,\omega^{\prime
})=|\omega\rangle\langle\omega^{\prime}|
\]
Then a generic relevant observable is $O_{R}\in\mathcal{O}_{R}$ (see
\cite{PRA2} eq. (42) or \cite{JPA}. eq. (42))%
\begin{equation}
O_{R}=|O_{R})=\int d\omega O(\omega)|\omega)+\int d\omega\int d\omega^{\prime
}O(\omega,\omega^{\prime})|\omega,\omega^{\prime}) \label{A}%
\end{equation}
and the generic states is (\cite{PRA2} eq. (45) or \cite{JPA}. eq. (45) )%
\begin{equation}
\rho_{R}=(\rho_{R}|=\int d\omega\rho(\omega)\widetilde{(\omega|}+\int
d\omega\int d\omega^{\prime}\rho(\omega,\omega^{\prime})\widetilde
{(\omega,\omega^{\prime}|} \label{B}%
\end{equation}
where $\widetilde{(\omega|},$ $\widetilde{(\omega,\omega^{\prime}|}$ will be
defined in eqs. (\ref{53'}) and (\ref{54}) but in the particular case $V=0$
(see also \cite{PRA2} eq. (44) or \cite{JPA}. eq. (45)). Then%
\[
\widetilde{(\omega|}O_{R})=O(\omega),\text{ }\widetilde{(\omega,\omega
^{\prime}|}O_{R})=O(\omega,\omega^{\prime})
\]
We will keep the treatment as general as possible, i.e. $O_{R}$ would be any
observable such that $O_{R}\in\mathcal{O}_{R}$\ and $\rho_{R}$ any state
$\rho_{R}\in\mathcal{O}_{R}^{\prime}$ \footnote{Namely, even more general than
the choice of EID $O_{R}=O_{S}\otimes I_{E}$ and more general than those of
other formalisms. This is why we can find the moving preferred basis in a
general case containing EID as particular case. Anyhow the analyticity
conditions must also be satisfied. In the case of EID we can substitute
$O_{R}$ by $O_{S}$ and $\rho_{R}(t)$ by $\rho_{S}(t)$ in some formulae e.g.
($\rho_{R}(t)|O_{R})=(\rho_{S}(t)|O_{S})$}. In fact, in the next subsection we
will only consider the generic mean value $(\rho_{R}(t)|O_{R})$ for two
paradigmatic model below. Model 1 with just one pole and the Khalfin effect
and Model 2 with two poles.

\subsection{Model 1. One pole and the Khalfin term:}

It can be proved (cf. (\cite{JPA}) eq. (67)) that the evolution equation of
the mean value $(\rho(t)|O)$ is
\begin{equation}
\langle O_{R}\rangle_{\rho(t)}=(\rho(t)|O_{R})=(\rho_{R}(t)|O_{R})=\int
_{0}^{\infty}\rho^{\ast}(\omega)O(\omega)\,d\omega+\int_{0}^{\infty}\int
_{0}^{\infty}\rho^{\ast}(\omega,\omega^{\prime})O(\omega,\omega^{\prime
})\,e^{i\frac{\omega-\omega^{\prime}}{\hbar}t}\,d\omega d\omega^{\prime}
\label{C}%
\end{equation}
i.e. this mean value in the case $V\neq0$ reads%

\begin{equation}
(\rho_{R}(t)|O_{R})=\int d\omega(\rho(0)|\Phi_{\omega})(\widetilde
{\Phi_{\omega}}|O_{R})+\int_{0}d\omega\int_{0}d\omega^{\prime}e^{\frac
{i}{\hbar}(\omega-\omega^{\prime})t}(\rho_{R}(0)|\Phi_{\omega\omega^{\prime}%
})(\widetilde{\Phi_{\omega\omega^{\prime}}}|O_{R})
\end{equation}

Where $O_{\omega}=(\widetilde{\Phi_{\omega}}|O_{R}),$ $O_{\omega\omega
^{\prime}}=(\widetilde{\Phi_{\omega\omega^{\prime}}}|O_{R}),$ $\rho_{\omega
}=(\rho_{R}(0)|\Phi_{\omega}),$ $\rho_{\omega\omega^{\prime}\text{ \ }}%
=(\rho_{R}(0)|\Phi_{\omega\omega^{\prime}}).$ These $\Phi$ vectors are defined
in eqs. (\ref{53'}) and (\ref{54}). Then, if we endow the functions with
analytical properties of subsection B and there is just one pole $z_{0}$ in
the lower halfplane, we can prove (\cite{JPA} eq. (70)) that%
\[
(\rho_{R}(t)|O_{R})=\int d\omega(\rho_{R}(0)|\Phi_{\omega})(\widetilde
{\Phi_{\omega}}|O_{R})+e^{\frac{i}{\hbar}(z_{0}^{\ast}-z_{0})t}(\rho
_{R}(0)|\Phi_{00})(\widetilde{\Phi_{00}}|O_{R})
\]%
\[
+\int_{\Gamma}dz^{\prime}e^{\frac{i}{\hbar}(z_{0}^{\ast}-z^{\prime})t}%
(\rho_{R}(0)|\Phi_{0z^{\prime}})(\widetilde{\Phi_{0z^{\prime}}}|O_{R}%
)+\int_{\Gamma^{\ast}}dze^{\frac{i}{\hbar}(z-z_{0})t}(\rho_{R}(0)|\Phi
_{0z})(\widetilde{\Phi_{0z}}|O_{R})
\]%
\begin{equation}
+\int_{\Gamma^{\ast}}dz\int_{\Gamma}dz^{\prime}e^{\frac{i}{\hbar}(z-z^{\prime
})t}(\rho_{R}(0)|\Phi_{zz^{\prime}})(\widetilde{\Phi_{zz^{\prime}}}|O_{R})
\label{70}%
\end{equation}
where $z_{0}$ $=\omega_{0}-\frac{i}{2}\gamma_{0},$ $\gamma_{0}>0$ and where
$|\Phi_{z}),(\widetilde{\Phi_{z}}|,|\Phi_{zz^{\prime}}),$ and $(\widetilde
{\Phi_{zz^{\prime}}}|$ are the analytical continuation in the lower half-plane
of (see (\cite{JPA} eq. (54))%
\begin{equation}
|\Phi_{\omega})=|\omega^{+}\rangle\langle\omega^{+}|,\text{ }\widetilde
{\text{(}\Phi_{\omega}|}=\widetilde{(\omega|},\text{ }|\Phi_{\omega
\omega^{\prime}})=|\omega^{+}\rangle\langle\omega^{+\prime}|, \label{53'}%
\end{equation}
and%
\begin{equation}
(\widetilde{\Phi_{\omega\omega^{\prime}}}|=\int d\varepsilon\lbrack
\langle\omega^{+}|\varepsilon\rangle\langle\varepsilon|\omega^{\prime+}%
\rangle-\delta(\omega-\varepsilon)\delta(\omega^{\prime}-\varepsilon
)](\widetilde{\varepsilon}|+\int d\varepsilon\int d\varepsilon^{\prime}%
\langle\omega^{+}|\varepsilon\rangle\langle\varepsilon^{\prime}|\omega
^{\prime+}\rangle\widetilde{(\varepsilon,\varepsilon^{\prime}}| \label{54}%
\end{equation}
and where $z_{0}$ is the simple pole of Figure 1 in the lower half-plane.
$|\Phi_{z}),(\widetilde{\Phi_{z}}|,|\Phi_{zz^{\prime}}),$ and $(\widetilde
{\Phi_{zz^{\prime}}}|$ can be defined as in the case of eq. (\ref{L.1}) and
(\ref{L.2}). The $|\Phi_{z}),(\widetilde{\Phi_{z}}|,|\Phi_{zz^{\prime}}),$ and
$(\widetilde{\Phi_{zz^{\prime}}}|$ can also be defined as a simple
generalization of the vectors $|f_{0}\rangle,$ $\langle\widetilde{f_{0}}|,$
$|f_{z}\rangle,$ and $\langle\widetilde{f_{z}}|$ (\cite{JPA}. eq. (42)). Then
the eqs. (\ref{53'}) and (\ref{54}) allow us to compute the limits
(\ref{INT-01}) and (\ref{INT-02'}) for any $\rho_{R}(0).$

Therefore we can conclude than the last four terms of equation (\ref{70})
\ vanish with characteristic times%
\begin{equation}
\frac{\hbar}{\gamma_{0}};\frac{2\hbar}{\gamma_{0}};\text{ }\frac{2\hbar
}{\gamma_{0}};\infty\label{TC}%
\end{equation}
respectively. Let us observe that

i-. The vanishing of the second, third, and fourth therms of eq. (\ref{70})
are \textit{exponential decaying}. This will also be the case in more
complicated models with many poles.

ii.- The $\infty$ means that the evolution of the last term of this equation
corresponds to a polynomial in $t^{-1\text{ }}$, \ i. e. to the
\textit{Khalfin evolution}. This is a very weak effect detected in 2006
\cite{KhalfinEx}. If there is a finite number of poles and the curve $\Gamma$
that yields below these poles the contribution of the integral along $\Gamma$
corresponds to the Khalfin effect \footnote{If the there is an infinite set of
poles at $z_{i},$ with imaginary part $-\frac{1}{2}\gamma_{i}$ such that
$lim_{i\rightarrow\infty}\gamma_{i}=\infty$, then we can choose a curve
$\Gamma_{j\text{ }}$below the poles a $\gamma_{1},\gamma_{2},...\gamma_{j}$.
Then the integral along the curve $\Gamma_{j}$ contains the effect of the
poles $\gamma_{j+1},\gamma_{j+2},...$and the Khalfin effect. Thus we can
choose the curve $\Gamma_{j\text{ }}$in such a way that the decaying times
corresponding to these poles, $t_{j+n}=\hbar/\gamma_{j+n}$ would be so small
that could be neglected.}. A closed system model for Khalfin effect can be
found in \cite{K1}, section 6, and an EID-like model in \cite{K2}, section 5.

Now for times $t>$ $t_{D}=\frac{\hbar}{\gamma_{0}}$, eq. (\ref{70}) reads%
\begin{equation}
(\rho_{R}(t)|O_{R})=\int d\omega(\rho(0)|\Phi_{\omega})(\widetilde
{\Phi_{\omega}}|O)+\int_{\Gamma^{\ast}}dz\int_{\Gamma}dz^{\prime}e^{\frac
{i}{\hbar}(z-z^{\prime})t}(\rho(0)|\Phi_{zz^{\prime}})(\widetilde
{\Phi_{zz^{\prime}}}|O) \label{700}%
\end{equation}
since for $t>$ $t_{D}=\frac{\hbar}{\gamma_{0}}$ the poles term has
vanished\footnote{Since $t_{D}$ is just an order of magnitude we consider that
the three first imaginary parts of eqs. (\ref{TC}) and (\ref{TC'}) are
essentially equivalent.}.

Let us diagonalize $\rho_{R}(t)$ of eq. (\ref{70}) as \footnote{Here, for the
sake of simplicity, we will use sum instead of integral, as we will do below
in all cases of diagonalization. Moreover, in many cases, the $O_{R}$ or the
initial conditions may just be expanded in a discrete basis of Hilbert space
(see below).}%
\begin{equation}
\rho_{R}(t)=\sum_{i}\rho_{i}(t)|i(t)\rangle\langle i(t)| \label{asterisco}%
\end{equation}
where \{$|i(t)\rangle\}$ is the moving eigenbasis of $\rho_{R}(t)$.

Then let us define a state $(\rho_{P}(t)|,$ the \textit{preferred state}, such
that, \textit{for all times,} it would be%
\begin{equation}
(\rho_{P}(t)|O_{R})=\int d\omega(\rho(0)|\Phi_{\omega})(\widetilde
{\Phi_{\omega}}|O)+\int_{\Gamma^{\ast}}dz\int_{\Gamma}dz^{\prime}e^{\frac
{i}{\hbar}(z-z^{\prime})t}(\rho(0)|\Phi_{zz^{\prime}})(\widetilde
{\Phi_{zz^{\prime}}}|O) \label{asterisco'}%
\end{equation}
So $\rho_{P}(t)$ is a state that evolves in a model with no poles and with
only the Khalfin term . These evolutions exist and can be found using an
adequate interaction \footnote{All these formulas are confirmed by the
coincidence of results with other methods: e.g. those used to study a
$^{208}Pb(2d_{5/2})$ proton state in a Woods-Saxon potential (see \cite{JPA}
Figure 3).}.%

\begin{figure}[t]
\centerline{\scalebox{0.7}{\includegraphics{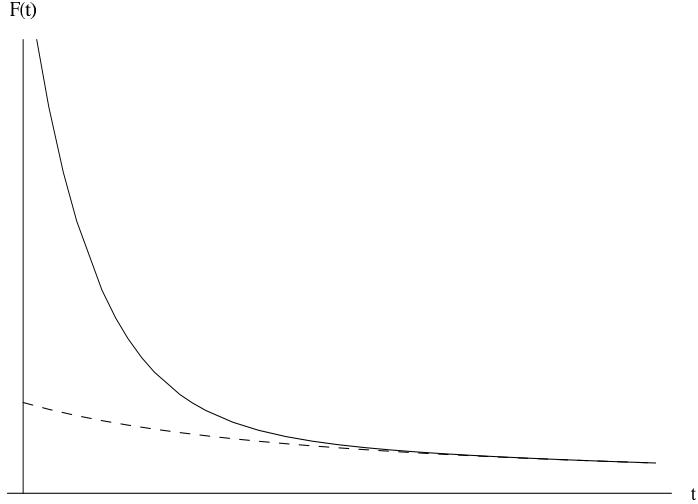}}} \vspace
*{0.cm}\caption{Evolution of $F(t)$ (solid line), $F_{Khalfin}(t)$ (dashed
line) and their coincidence limit at $t_{D}$.}%
\label{fig 2}%
\end{figure}

It is quite clear that for $t>t_{D}$ $\rho_{R}(t)\neq$ $\rho_{P}(t)$ while for
$t<t_{D}$ $\rho_{R}(t)=$ $\rho_{P}(t)$ and that for $t\rightarrow t_{D}$
$\rho_{R}(t)\rightarrow$ $\rho_{P}(t)$ and also all their derivatives.

The eigen states of the $\rho_{P}(t)$ are those that we will choose for the
moving decoherence basis. In fact, diagonalizing $\rho_{P}(t)$ we have
\begin{equation}
\rho_{P}(t)=\sum_{j}\rho_{j}(t)\widetilde{|j(t)\rangle}\widetilde{\langle
j(t)|} \label{asterisco'-b}%
\end{equation}
and when $t\rightarrow$ $t_{D}=\frac{\hbar}{\gamma_{0}}$ we have that
$\rho(t)\rightarrow\rho_{P}(t)$ so from eqs. (\ref{asterisco}) and
(\ref{asterisco'-b}) we see that the eigenbasis of $\rho(t)$ and $\rho_{P}(t)$
also converge
\begin{equation}
\left\{  |i(t)\rangle\right\}  \rightarrow\{\widetilde{|j(t)\rangle}\}
\label{asterisco'-c}%
\end{equation}
Namely the basis $\left\{  |i(t)\rangle\right\}  $ converge to $\{\widetilde
{|j(t)\rangle}\}$ and therefore $\rho_{R}(t)$ becomes diagonal in
$\{\widetilde{|j(t)\rangle}\}$. Thus $\{\widetilde{|j(t)\rangle}\}$ is our
definition for the \textit{moving preferred basis} for this case. Since
$\rho_{R}(t)$ becomes diagonal in the just defined preferred basis
$\{\widetilde{|j(t)\rangle}\}$ when $t\rightarrow t_{D}$ and $t_{D}%
=\frac{\hbar}{\gamma_{0}}$ is really the definition of the decoherence time.
In this model the relaxation time $t_{R}$ is the corresponding to the Khalfin
term, i.e. an extremely long time so%
\begin{equation}
t_{D}\ll t_{R} \label{nuevo-0}%
\end{equation}

\subsection{Model 2: Two poles are considered and the Khalfin term is
neglected.}

The Khalfin term is so small (see \cite{KhalfinEx}) that can be neglected in
most of the experimental cases. So let us consider the case of two poles
$z_{0}$ and $z_{1}$ (and no relevant Khalfin term) where eq. (\ref{70}) reads:%

\[
(\rho_{R}(t)|O_{R})=\int d\omega(\rho_{R}(0)|\Phi_{\omega})(\widetilde
{\Phi_{\omega}}|O_{R})+e^{\frac{i}{\hbar}(z_{0}^{\ast}-z_{0})t}(\rho
_{R}(0)|\Phi_{00})(\widetilde{\Phi_{00}}|O_{R})+
\]%
\begin{equation}
e^{\frac{i}{\hbar}(z_{1}^{\ast}-z_{0})t}(\rho_{R}(0)|\Phi_{10})(\widetilde
{\Phi_{10}}|O_{R})+e^{\frac{i}{\hbar}(z_{0}^{\ast}-z_{1})t}(\rho_{R}%
(0)|\Phi_{01})(\widetilde{\Phi_{01}}|O_{R})+e^{\frac{i}{\hbar}(z_{1}^{\ast
}-z_{0})t}(\rho_{R}(0)|\Phi_{11})(\widetilde{\Phi_{11}}|O_{R}) \label{70'}%
\end{equation}
where $z_{0}$ $=\omega_{0}-\frac{i}{2}\gamma_{0},$ $\gamma_{0}>0$ , $z_{1}$
$=\omega_{1}-\frac{i}{2}\gamma_{1},$ $\gamma_{1}>0,$ and we will also consider
that $\gamma_{0}\ll\gamma_{1}$ (see \cite{Manoloetal} section 3, for details).
Then the four characteristic times (\ref{TC}) now read.%
\begin{equation}
\frac{\hbar}{\gamma_{0}};\frac{\hbar}{\gamma_{1}+\gamma_{0}};\frac{\hbar
}{\gamma_{1}+\gamma_{0}}\approx\frac{\hbar}{\gamma_{1}} \label{TC'}%
\end{equation}
Now for times $t>$ $t_{D}=\frac{\hbar}{\gamma_{1}}$, eq. (\ref{700}) reads%
\[
(\rho_{R}(t)|O_{R})=\int d\omega(\rho_{R}(0)|\Phi_{\omega})(\widetilde
{\Phi_{\omega}}|O_{R})+e^{\frac{i}{\hbar}(z_{0}^{\ast}-z_{0})t}(\rho
_{R}(0)|\Phi_{00})(\widetilde{\Phi_{00}}|O_{R})
\]
and we can define a state $(\rho_{P}(t)|$ such that, for \textit{all times},
it would be%
\begin{equation}
(\rho_{P}(t)|O_{R})=\int d\omega(\rho_{R}(0)|\Phi_{\omega})(\widetilde
{\Phi_{\omega}}|O_{R})+e^{\frac{i}{\hbar}(z_{0}^{\ast}-z_{0})t}(\rho
_{R}(0)|\Phi_{00})(\widetilde{\Phi_{00}}|O_{R}) \label{asterisco''}%
\end{equation}
Repeating the reasoning of eqs. (\ref{700}) to (\ref{asterisco'-c}) we can see
that, diagonalizing this last equation, as in eq. (\ref{asteriscio'-b}), we
obtain the moving preferred basis. Then in this case we see that the
\ relaxation is obtained by an exponential dumping (not a Khalfin term) and
\begin{equation}
t_{R}=\frac{\hbar}{\gamma_{0}}\gg t_{D}=\frac{\hbar}{\gamma_{1}} \label{nuevo}%
\end{equation}
Again, in this case when $t\rightarrow$ $t_{D}=\frac{\hbar}{\gamma_{0}}$ we
have that $\rho_{R}(t)\rightarrow\rho_{P}(t)$ so once more we reach eq.
(\ref{asterisco'-c}). Namely $\rho(t)$ becomes diagonal in the moving
preferred basis in a time $t_{D}$.

\subsubsection{Remark}

Before considering the many poles case let us make some general remarks.

i.- Let us observe that some $(\widetilde{\Phi_{\omega}}|O_{R}),$
$(\widetilde{\Phi_{0z^{\prime}}}|O_{R}),$ $(\widetilde{\Phi_{0z}}|O_{R})$ and
$(\widetilde{\Phi_{zz^{\prime}}}|O_{R})$ may be zero, depending in the
observable $O_{R},$ so, in the case of many poles, may be some poles can be
detected by $O_{R}$ and others may not be detected and disappear from the
formulae (see Appendix).

This also is the cases for the initial conditions: $(\rho_{R}(0)|\Phi_{\omega
}),$ $(\rho_{R}(0)|\Phi_{0z^{\prime}}),$ $(\rho_{R}(0)|\Phi_{0z}),$ and
$(\rho_{R}(0)|\Phi_{zz^{\prime}})$ may be zero. But also the $O_{R}$ or the
$\rho_{R}(0)$ may create some poles. So some poles may be eliminated or
created by the observables or the initial conditions while others may be
retained. But in general we will choose $O_{R}$ and $\rho_{R}(0)$ in such a
way that they would neither create or eliminate poles.

ii.- From what we have learned in both models (see eqs. (\ref{nuevo-0}) and
(\ref{nuevo})) we always have%
\begin{equation}
t_{D}<t_{R} \label{ECR-01}%
\end{equation}

\section{The general case}

\subsection{Relaxation, decoherence, and moving preferred basis.}

Let us now consider the general case of a system with $N+1$ poles at
$z_{i}=\omega_{i}^{\prime}-i\gamma_{i}.$ These poles are the ones that remain
after $O_{R}$ and $\rho_{R}(0)$ have eliminated (or created) some poles (see
remark above). In this case it is easy to see that eq. (\ref{70'}) (with no
Khalfin term) becomes:%
\begin{equation}
(\rho_{R}(t)|O_{R})=(\rho_{R\ast}|O_{R})+a_{0}(t)\exp\left(  -\frac{\gamma
_{0}}{\hbar}t\right)  +\sum_{i=1}^{N}a_{i}(t)\exp\left(  -\frac{\gamma_{i}%
}{\hbar}t\right)  =(\rho_{R\ast}|O_{R})+a_{0}(t)\exp\left(  -\frac{\gamma_{0}%
}{\hbar}t\right)  +f(t) \label{Suma}%
\end{equation}
where $(\rho_{R\ast}|O_{R})$ is the final equilibrium value of $(\rho
_{R}(t)|O_{R})$ and the $a_{i}(t)$ are real oscillating functions$.$ In the
most general case the $z_{i}$ will be placed either at random or not. Anyhow
in both cases they can be ordered as\footnote{For simplicity we will consider
only the case $\gamma_{0}<\gamma_{1}<\gamma_{2}<...$Other special cases will
be consider elsewhere.}%
\[
\gamma_{0}\leq\gamma_{1}\leq\gamma_{2}\leq...
\]
So we have plotted $F(t)=(\rho_{R}(t)|O_{R})-(\rho_{R\ast}|O_{R})$ in figure 3%

\begin{figure}[t]
\centerline{\scalebox{0.7}{\includegraphics{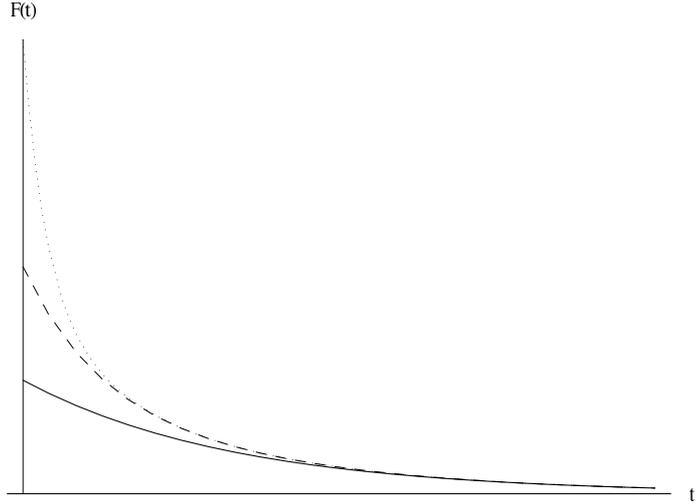}}} \vspace
*{0.cm}\caption{Evolution of $F(t)$ (solid line), $F_{\gamma_{0},\gamma_{1}%
}(t)$ (dashed line), $F_{\gamma_{0}}(t)$ (dot line)\ and their coincidence
limit at $t_{D}$. We can see the dominant components in different periods of
time.}%
\label{fig 3}%
\end{figure}

Then if $\gamma_{0}<\gamma_{1}$ it is quite clear that the relaxation time is
$t_{R}=\frac{\hbar}{\gamma_{0}}.$ So the relaxation time is defined with no ambiguity

Let us now consider the decoherence time. Really each pole $z_{i}$ defines a
decaying mode with characteristic time%
\[
t_{i}=\frac{\hbar}{\gamma_{i}}%
\]
Moreover the poles contain the essence of the decaying phenomenon and the
definition of the decoherence time depends on their distribution and other
data like the initial condition \footnote{In a completely random distribution
of poles , the best choice seem to be%
\[
t_{R}=\frac{\hbar}{\gamma_{0}},\text{ \ }t_{D}=\frac{\hbar}{\gamma_{1}}%
\]
Then, in this case
\begin{equation}
(\rho_{P}(t)|O_{R})=(\rho_{R\ast}|O_{R})+a_{0}(t)\exp\left(  -\frac{\gamma
_{0}}{\hbar}t\right)  \label{def}%
\end{equation}
and the moving preferred basis would be $\{\widetilde{|j(t)\rangle}\},$ i.e.
the basis that diagonalizes $\rho_{P}(t),$ as before. But, as we will see, we
can improve this definition with a more general one that it is valid in a
completely general case.}. In fact our experience, obtained by the study of
many models of the literature, is that, $t_{R}$ seems to be given by the
slowest mode irrespectively of the initial conditions. Then we generally
define the relaxation time as
\[
t_{R}=\frac{\hbar}{\gamma_{0}}%
\]
In doing so we have used the first recipe of section II. to use the pole
nearest to the real axis. This $t_{R}$ will coincide with the one of the model
of the next section.

But the initial conditions seem essential for the definition \ of $t_{D}$. In
fact the decoherence time is related to macroscopicity and macroscopicity is
defined by the initial conditions.

Now, to introduce the initial condition, let us define:
\[
f(t)=\sum_{i=1}^{N}a_{i}(t)\exp\left(  -\frac{\gamma_{i}}{\hbar}t\right)
,\text{ \ }f^{\prime}(t)=-\frac{i}{\hbar}\sum_{i=1}^{N}a_{i}(t)\gamma_{i}%
\exp\left(  -\frac{\gamma_{i}}{\hbar}t\right)
\]
so at $t=0$ we can write the initial conditions as%
\[
f(0)=\sum_{i=1}^{N}a_{i}(0),\text{ \ \ \ \ \ \ \ \ \ }f^{\prime}(0)=-\frac
{i}{\hbar}\sum_{i=1}^{N}a_{i}(0)\gamma_{i}%
\]
Let us call $f(t)=const.\exp g(t)\sim\exp g(t)$, and let us make a Taylor
expansion of $g(t)$ as%
\begin{equation}
g(t)=g(0)+g^{\prime}(0)t+\frac{1}{2}g^{\prime\prime}(0)t^{2}+... \label{app}%
\end{equation}
Let us now adimensionalize this expression introducing the adimensional
variable $\eta=\frac{t}{t_{R}}$ where we have used $t_{R}$ the only
characteristic time we have up to now.%
\begin{equation}
g(\eta)=g(0)+g^{\prime}(0)t_{R}\eta+\frac{1}{2}g^{\prime\prime}(0)t_{R}%
^{2}\eta^{2}+...
\end{equation}
where $g(t)=g(\eta),$ $g^{\prime}(0)t_{R},$ $g^{\prime\prime}(0)t_{R}^{2},...$
are adimensional. Now, as we explained in the introduction, the essential
challenge of the second period of the history of decoherence was to obtain a
decoherence time clearly smaller than the relaxation time. So let us postulate
that the decoherence time is $t_{D}\ll t_{R}$. i.e. $\eta\ll1.$ Then with this
condition we have the approximations:%
\begin{equation}
g(\eta)=g(0)+g^{\prime}(0)t_{R}\eta\text{ \ \ \ or \ \ }g(t)=g(0)+g^{\prime
}(0)t \label{tesis}%
\end{equation}
where%
\[
g(0)=\log f(0)=\log\sum_{i=0}^{N}a_{i}(0),\text{ \ ..}g\prime(0)=\frac
{f^{\prime}(0)}{f(0)}=-\frac{1}{\hbar}\frac{\sum_{i=0}^{N}a_{i}\gamma_{i}%
}{\sum_{i=0}^{N}a_{i}}%
\]
These equations contain the initial conditions. Then in this approximation:
\begin{equation}
f(t)=\exp g(0)\exp tg^{\prime}(0)=f(0)\exp\left(  -\frac{t}{\hbar}\frac
{\sum_{i=0}^{N}a_{i}\gamma_{i}}{\sum_{i=0}^{N}a_{i}}\right)  =f(0)\exp\left(
-\frac{\gamma_{eff}}{\hbar}t\right)  \label{Omneslike}%
\end{equation}
where%
\begin{equation}
\gamma_{eff}=\frac{\sum_{i=0}^{N}a_{i}\gamma_{i}}{\sum_{i=0}^{N}a_{i}}\text{
\ \ \ and \ \ }f(t)=f(0)\exp\left(  -\gamma_{eff}\frac{t}{\hbar}\right)
\label{eff}%
\end{equation}
Then as $\gamma_{eff}>\gamma_{0}$ for \ the period in study $t_{D}<t_{R}$ we
have obtain a typical decaying very fast evolution that we consider the one
that produce the decoherence phenomenon. So we define
\begin{equation}
t_{D}=\frac{\hbar}{\gamma_{eff}} \label{34'}%
\end{equation}
Then $\gamma_{eff}$ and $t_{D}$ are both functions of the initial conditions.
We will see that this $t_{D}$ coincide with the one of the Omn\`{e}s example
in the next subsection.

Let us now go to the definition of the moving preferred basis. It is clear
that, for the time $t>t_{D},$ the modes with characteristic times $t_{i}%
<t_{D}$ (i.e. $\gamma_{i}>\gamma_{eff}$), that we will call the \textit{fast
modes}, have become negligible in eq. (\ref{Suma}) and therefore in this case
$\rho_{P}(t)$ must defined as%

\begin{equation}
(\rho_{P}(t)|O_{R})=(\rho_{R\ast}|O_{R})+\sum_{i=0}^{M}a_{i}(t)\exp\left(
-\frac{\gamma_{i}}{\hbar}t\right)  \label{EV}%
\end{equation}
where the sum in this equation contains only the $M<N$ poles such that
$\gamma_{i}<\gamma_{eff}$, that correspond to the \textit{slow modes}
\footnote{E. g. in the case of the eq. (\ref{asterisco''}) only the
$z_{0\text{ }}$appears in the r.h.s. of the equation. This is the result of
eq. (\ref{EV}), in the case $a_{0}=a_{1},$ and $\gamma_{0}<\gamma_{1}.$}. This
is our \textit{adiabatic} choice since we have chose the slow modes of
decaying to define $\rho_{P}(t)$ and rejected the fast modes\footnote{The
requirement of macroscopicity introduced the initial conditions in the play.
In turn these initial conditions define the $\gamma_{eff}$ and then define the
fast and slow modes. Our adiabatic choice corresponds to keep the slow modes
and disregard the fast ones. Thus for us the \textit{robust} modes are the
slow ones, since they are the less affected by the interaction with the
environment, that creates the poles, if compared with the fast ones. This is
our notion of \textit{robustness. \ }Analogously, if we compute the
\textit{linear entropy} we will have slower variation of this entropy if we
consider only the slow modes that if we consider all the modes (including the
fast ones). This would be our minimization of the linear entropy: the moving
preferred basis evolution contains only the slow modes.}. Moreover, when
$t>t_{D}$ the motions produce by the fast modes, such that $\gamma_{i}%
>\gamma_{eff},$ \ namely those with motions faster than the one of the
evolution of eq. (\ref{eff}$_{2}$), are ceased to be relevant for $\rho
_{R}(t)$ and $\rho_{P}(t)\rightarrow\rho_{R}(t).$ Then we call diagonalize
$\rho_{P}(t)$ and we obtain the moving preferred basis $\{\widetilde
{|j(t)\rangle}\}$, the eigen basis of $\rho_{P}(t)$, that evolves only
influenced by the poles such that $\gamma_{i}<\gamma_{eff}$, and such that,
when $t\rightarrow t_{D},$ $\{\widetilde{|j(t)\rangle}\}\rightarrow$ the
eiegenbasis of $\rho_{R}(t).$ This $\{\widetilde{|j(t)\rangle}\}$ is our
candidate for a general definition of moving preferred basis. Finally if
$\gamma_{0}<\gamma_{1}<\gamma_{2}<...$we have that%
\[
t_{R}<t_{D}%
\]

\subsection{\label{Polos}The Omn\`{e}s or Lee-Friedrich model..}

Our more complete and simplest example of decoherence in open systems is the
Omn\`{e}s \textquotedblleft pendulum\textquotedblright\ (i. e. oscillator) in
a bath of oscillators, that we will compare with the poles theory in the
following subsections. In fact the Omn\`{e}s model could be considered a poles
model if we retain the poles and neglect the Khalfin term. Moreover in the
Omn\`{e}s philosophy the moving preferred basis must be related to some
\textquotedblleft collective variables\textquotedblright\ in such a way that
they would be experimentally accessible. In this case this variable is the
center of mass of the pendulum, i. e. the mean value of the position of a
coherent state. In \cite{Omnesazul} page 285 a one dimensional "pendulum" (the
system) in a bath of oscillators (the environment) is considered. The
Hamiltonian reads \footnote{This Hamiltonian is similar to the one of equation
(\ref{Hamil}) and equation (\ref{PRE}). In fact, in some stages of the
treatment Omn\`{e}s is forced to go to the continuos spectrum. A complete
treatment of this continuous model can be found in \cite{PRE}. The present of
the factor $\hbar$ in the Hamiltonian will change some formulas (as
$t_{R}=\hbar/\gamma\rightarrow t_{R}=1/\gamma)$ but we prefer to follow the
Omn\`{e}s formalism in this section}%
\begin{equation}
H=\hbar\omega a^{\dagger}a+\sum_{k}\hbar\omega_{k}b_{k}^{\dagger}b_{k}%
+\hbar\sum_{k}(\lambda_{k}a^{\dagger}b_{k}+\lambda_{k}^{\ast}ab_{k}^{\dagger})
\label{1}%
\end{equation}
where $a^{\dagger}(a)$ is the creation (annihilation) operator for the system,
$b_{k}^{\dagger}(b_{k})$ are the creation (annihilation) operator for each
mode of the environment, $\omega$ and $\omega_{k}$ are the energies of the
system and each mode of the environment and $\lambda_{k}$ are the interaction coefficients.

Then let consider a state%
\[
|\psi(t)\rangle=a|\alpha_{1}(t)\rangle\prod_{k}|\beta_{k1}(t)\rangle
+b||\alpha_{2}(t)\rangle\prod_{k}|\beta_{k2}(t)\rangle
\]
where $|\alpha_{1}(0)\rangle,|\alpha_{2}(0)\rangle$ are \textit{coherent}
states for the "system" corresponding to the operator $a^{\dagger}$ and
$|\beta_{k1}(0)\rangle,$ $|\beta_{k2}(0)\rangle$ are a coherent state for the
environment corresponding to the operator $b_{k}^{\dagger}.$ Let the initial
condition be%
\begin{equation}
|\psi(0)\rangle=a|\alpha_{1}(0)\text{ \{}\beta_{k1}(0)=0\}\rangle+b|\alpha
_{2}(t),\{\beta_{k2}(0)=0\rangle\label{in}%
\end{equation}
Then
\begin{equation}
\rho_{R}(0)=Tr_{E}|\psi(0)\rangle\langle\psi(0)|\text{ and }\rho_{R}%
(t)=Tr_{E}|\psi(t)\rangle\langle\psi(t)| \label{tr}%
\end{equation}
Moreover Omn\`{e}s shows that, under reasonable hypotheses and approximations
(that correspond to the elimination of the Khalfin terms, see below), the
evolution of the $|\alpha_{1}(t)\rangle,|\alpha_{2}(t)\rangle$ is given by
\begin{equation}
\alpha(t)=\alpha(0)\exp[-i(\omega+\delta\omega)t-\gamma t]+\text{small
fluctuations} \label{RT-01}%
\end{equation}
where $\delta\omega$ is a shift and $\gamma$\ a dumping coefficient that
produces that the system would arrive at a state of equilibrium at
$t_{R}=1/\gamma$\footnote{So Omn\`{e}s $\gamma$ corresponds to our $\gamma
_{0}.$}, the \textit{relaxation time} of the system, (the small fluctuations
are usually neglected)

In the next subsections using the concepts of the previous sections we will
prove that the Omn\`{e}s model is a particular case of our general scheme of
the section III.A. Let us now consider the condition of \textit{experimentally
accessibility. }In fact, in the model under consideration, the initial states
corresponds to the linear combination of two coherent, macroscopically
different states $|\alpha_{1}(0)\rangle,|\alpha_{2}(0)\rangle$ that evolve to
$|\alpha_{1}(t)\rangle,|\alpha_{2}(t)\rangle$ .

Now the diagonal part of $\rho_{R}(t)$ reads
\[
\rho_{R}^{(D)}(t)=|a|^{2}|\alpha_{1}(t)\rangle\langle\alpha_{1}(t)|+|b|^{2}%
|\alpha_{2}(t)\rangle\langle\alpha_{2}(t)|
\]
and, it can easily be shown \cite{Omnesazul} that, with the choice of initial
conditions of eqs. (\ref{CI-17}) and (\ref{CI-18}), that the non diagonal part
of $\rho_{R}(t)$ is%

\begin{equation}
\rho_{R}^{(ND)}(t)=(ab^{\ast}|\alpha_{1}(0)\rangle\langle\alpha_{2}%
(0)|+ba^{\ast}|\alpha_{2}(0)\rangle\langle\alpha_{1}(0)|)\exp\left[  -\frac
{1}{2}\frac{m\omega}{\hbar}(x_{1}(0)-x_{2}(0))^{2}(1-e^{-\gamma t})\right]
\label{131}%
\end{equation}
Then if $t\ll t_{R}=\frac{1}{\gamma}$ (that will be the case if $L_{0}%
=|x_{2}(0)-x_{1}(0)|$ is very big) we have%
\begin{equation}
\rho_{R}^{(ND)}(t)\sim(ab^{\ast}|\alpha_{1}(0)\rangle\langle\alpha
_{2}(0)|+ba^{\ast}|\alpha_{2}(0)\rangle\langle\alpha_{1}(0)|)\exp\left[
-\frac{1}{2}\frac{m\omega}{\hbar}(x_{1}(0)-x_{2}(0))^{2}(1-1+\frac{t}{t_{R}%
}+...)\right]  \label{OM}%
\end{equation}
where $x_{1}(0),x_{2}(0)$ are the initial mean value of the position of the
Wigner transform of the two coherent states $|\alpha_{1}(t)\rangle
\langle\alpha_{1}(t)|,|\alpha_{2}(t)\rangle\langle\alpha_{2}(t)|$. This
decaying structure is obviously produced by the combination of the initial
states and the particular evolution of the system according to the discussion
in the introduction of the this section. Then, since $\rho_{R}^{(ND)}%
(t)\rightarrow0$ when $t\rightarrow\infty,$ $\rho_{R}(t)$ decoheres in the
decoherence basis \{$|\alpha_{1}(t)\rangle,|\alpha_{2}(t)\rangle\}$, which is
the moving preferred basis, and the decoherence time of the system is%

\begin{equation}
t_{D}(L_{0})\sim\lbrack m\omega(x_{1}(0)-x_{2}(0))^{2}]^{-1}t_{R} \label{131'}%
\end{equation}
where $L_{0}=$%
$\vert$%
$x_{1}(0)-x_{2}(0)|.$

In the next subsection we will see that we are dealing with a many poles model
where the effect of decoherence is produced by these poles and the particular
coherent states initial conditions, which produce a "new collective pole mode"
with $\gamma_{eff}=$ $\frac{1}{2}\frac{m\omega^{2}}{\hbar}(x_{1}%
(0)-x_{2}(0))^{2}$

In the case of the "pendulum" the moving preferred basis \{$|\alpha
_{1}(t)\rangle,|\alpha_{2}(t)\rangle\}$ is clear experimentally accessible
since, in principle, the mean value of the position $x_{1}(t),x_{2}(t),$ of
the two coherent states $|\alpha_{1}(t)\rangle,|\alpha_{2}(t)\rangle$ can be
measured and the $x_{1}(0)$ and $x_{2}(0)$ turn out to be two "collective
variables" (since they are mean values). In fact, in this formalism, the main
characteristic of the moving preferred basis is to be related to the
"collective variables". Moreover the decoherence time $t_{D}$ \ depends on the
initial distance $L_{0}=|x_{1}(0)-x_{2}(0)|$ so we can have different
decoherence times depending on the initial conditions.

Let us now consider that
\[
\langle\alpha_{1}(t)|\alpha_{2}(t)\rangle=\exp\left[  -\frac{|\alpha
_{1}-\alpha_{2}|^{2}}{2}+i\frac{\Phi}{2}\right]  ,\text{ \ \ \ }%
\Phi=\operatorname{Im}(\alpha_{1}\alpha_{2}^{\ast}-\alpha_{1}^{\ast}\alpha
_{2})
\]
where
\[
|\alpha_{1}-\alpha_{2}|=(2m\hbar\omega)^{-\frac{1}{2}}[m^{2}\omega^{2}%
(x_{1}(t)-x_{2}(t))^{2}+(p_{1}(t)-p_{2}(t))^{2}]^{\frac{1}{2}}%
\]
So:

i.- $\langle\alpha_{1}(t)|\alpha_{1}(t)\rangle=1$ even if in general
$\langle\alpha_{1}(t)|\alpha_{2}(t)\rangle\neq0$.

ii.- When $(x_{1}(0)-x_{2}(0))^{2}\rightarrow\infty$ or $(p_{1}(t)-p_{2}%
(t))^{2}\rightarrow\infty$ we have $\langle\alpha_{1}(t)|\alpha_{2}%
(t)\rangle\rightarrow0.$

Thus when the distance between the two centers of the coherent states is very
big we have a small $t_{D}$ and the basis \{$|\alpha_{1}(t)\rangle,|\alpha
_{2}(t)\rangle\}$would be almost orthonormal. These are the main
characteristics of the experimental accessible decoherence basis of Omn\`{e}s.

But it is important to insist that, generally, $\left\{  \left\vert \alpha
_{i}(t)\right\rangle \right\}  $ is only a \textit{non-orthonormal moving
preferred basis}, that we can approximately suppose orthonormal only in the
macroscopic case, that is to say, when $x_{1}(0)$ and $x_{2}(0)$ are far apart.

In conclusion, in this macroscopic case $\left\{  \left\vert \alpha
_{i}(t)\right\rangle \right\}  $ becomes a orthonormal \textit{ moving
preferred basis }where $\rho_{R}(t)$ becomes diagonal in a very small time.
This will be the case of the decoherence basis in \cite{OmnesRojo}, chapter
17, and in many examples that we can find in the bibliography
(\cite{Paz-Zurek}, \cite{Paz-Habib-Zurek}, \cite{Ex}). Without this
macroscopic property it is difficult to find any trace of a Boolean logic in
the moving decoherence basis context of the general case or in this section.
In fact, Omn\`{e}s obtains the Boolean logic by a complete different way (see
chapter 6 of \cite{Omnesazul}).

Anyhow in this particular model the moving preferred basis has a perfect
example for the macroscopic case. Let us now present the relation of this
formalism with the poles theory.

\subsection{\label{PolosH} The Lee-Friedrich model. The relaxation time.}

Particular important models can be studied, like the one in \cite{PRE}, with
Hamiltonian\footnote{The introduction of $\hbar$ in the equation below
produces some changes in the dimensions of some variables, but we prefer to
use the Omn\`{e}s convention in this sub-section also to facilitate the
comparison.}%

\begin{equation}
H=\hbar\omega_{0}a^{\dagger}a+\int\hbar\omega_{\mathbf{k}}b_{\mathbf{k}%
}^{\dagger}b_{\mathbf{k}}d\mathbf{k}+\hbar\int\lambda_{\mathit{k}}(a^{\dagger
}b_{\mathbf{k}}+ab_{\mathbf{k}}^{\dagger})d\mathbf{k} \label{PRE}%
\end{equation}

i.e. a continuous version of (\ref{1}). In this continuous version we are
forced to endow the scalar $(\rho_{R}(t)|O_{R})$ with the some analyticity
conditions. Precisely function $\lambda_{\mathit{k}}$ (where $k=\omega
_{k}=|\mathbf{k|)}$ is chosen in such a way that%
\begin{equation}
\eta_{\pm}(\omega_{k})=\omega_{k}-\omega_{0}-\int\frac{d\mathbf{k}%
\lambda_{\mathit{k}}^{2}}{\omega_{k}-\omega_{k^{\prime}}\pm i0}
\label{cuac-00}%
\end{equation}
which does not vanish for $k\in\mathbb{R}^{+},$ and its analytic extension
$\eta_{+}(z)$ to the lower half plane only has a simple pole at $z_{0}$. This
fact will have influence on the poles of $(\rho_{R}(t)|O_{R})$ as in section
II and we know that the study of $(\rho_{R}(t)|O_{R})$ is the essential way to
understand the whole problem (see section I A).

The Hamiltonian (\ref{PRE}) is sometimes called the Lee-Friedrich Hamiltonian
and it is characterized by the fact that it contains different \textit{number
of modes sector} (number of particle sectors in QFT). In fact, $a^{\dagger}$
and $b_{\mathbf{k}}^{\dagger}$ are creation operators that allow to define
these numbers of mode sectors. e. g. the one mode sector will contain states
like $a^{\dagger}|0\rangle$ and $b_{\mathbf{k}}^{\dagger}|0\rangle$ (where
$a|0\rangle=$ $b_{\mathbf{k}}|0\rangle=0).$ Then the action of $\exp\left(
-\frac{1}{\hbar}Ht\right)  $ (or simple the one of $H)$ will conserve the
number of modes of this sector in just one mode, since in \ (\ref{PRE}) all
the destruction operators are preceded by a creation operator. This also is
the case for the $n-$mode sector. The Hamiltonian of the one mode sector, is
just the one of the so called Friedrich model i. e.
\begin{equation}
H_{F}=\hbar\omega_{0}\left\vert 1\right\rangle \left\langle 1\right\vert
+\int\hbar\omega_{k}\left\vert \omega\right\rangle \left\langle \omega
\right\vert d\omega+\hbar\int\left(  \lambda(\omega)\left\vert \omega
\right\rangle \left\langle 1\right\vert +\lambda^{\ast}(\omega)\left\vert
1\right\rangle \left\langle \omega\right\vert \right)  d\omega\label{cuac-04}%
\end{equation}
(expressed just in variable $\omega,$ the one that will be analytically
continued$).$ As a consequence of the analyticity condition above this simple
Friedrich model shows just one resonance. In fact, this resonance is produced
in $z_{0}$. Let $H_{F}$ be the Hamiltonian of the complex extended Friedrich
model, then\footnote{Only symbolically, since the poles belong to the scalar
$(\rho(t)|O),$ as in section II.} :%
\begin{equation}
H_{F}|z_{0}\rangle=z_{0}|z_{0}\rangle,\text{ \ \ \ }H_{F}|z\rangle
=z|z\rangle\label{cuac-05}%
\end{equation}
where $z_{0}=\omega_{0}+\delta\omega_{o}-i\gamma_{0}=\omega_{0}^{\prime
}-i\gamma_{0}$ is the only pole and $z\in\Gamma$.

The Lee-Friedrich model, describing the interaction between a quantum
oscillator and a scalar field, is extensively analyzed in the literature.
Generally, this model is studied by analyzing first the one excited mode
sector, i.e. the Friedrich model. Then, if we compute the pole, of this last
model, up to the second order in $\lambda_{k}$ we obtain that
\begin{equation}
z_{0}=\omega_{0}+\int\frac{d\mathbf{k}^{\prime}\lambda_{\mathit{k}^{\prime}%
}^{2}}{\omega_{0}-\omega_{k}+i0} \label{cuac-01}%
\end{equation}

So the pole (that will corresponds to the pole closest to the real axis in the
Lee-Friedrich model) can be calculated (see \cite{LCIB} eq. (42)). These
results coincide (mutatis mutandis) with the one of Omn\`{e}s book
\cite{Omnesazul}\ page 288, for the relaxation time. In fact:
\begin{equation}
\frac{1}{\omega_{0}-\omega^{\prime}+i0}=P\left(  \frac{1}{\omega_{0}%
-\omega^{\prime}}\right)  -i\pi\delta(\omega_{0}-\omega^{\prime})
\label{cuac-02}%
\end{equation}
where $P$ symbolizes the \textquotedblleft principal part\textquotedblright,
so%
\begin{equation}
z_{0}=\omega_{0}+P\int\frac{d\mathbf{k}^{\prime}\lambda_{\mathbf{k}^{\prime}%
}^{2}}{\omega_{0}-\omega_{k}}-i\pi\int d\mathbf{k}^{\prime}\lambda
_{\mathbf{k}^{\prime}}^{2}\delta(\omega_{0}-\omega_{k}) \label{cuac-03}%
\end{equation}
Then if $d\mathbf{k}=n(\omega)d\omega$ we have%
\begin{equation}
\delta\omega_{0}=P\int\frac{n(\omega^{\prime})d\omega^{\prime}\lambda
_{\omega^{\prime}}^{2}}{\omega_{0}-\omega^{\prime}},\text{ \ \ \ \ }\gamma
_{0}=\pi\int n(\omega^{\prime})d\omega^{\prime}\lambda_{\omega^{\prime}}%
^{2}\delta(\omega_{0}-\omega^{\prime}) \label{Omnes}%
\end{equation}
namely the results of \cite{Omnesazul}\ page 288, and the one contained in eq.
(\ref{RT-01}).:%
\begin{equation}
z_{0}=\left(  \omega_{0}+\delta\omega_{0}\right)  -i\gamma_{0}=\omega
_{0}^{\prime}-i\gamma_{0} \label{cuac}%
\end{equation}
So the Omn\`{e}s result for the relaxation time \textit{coincides}, as we have
already said, with the one obtained by the pole theory, precisely
\[
t_{R}=\frac{1}{\gamma_{0}}%
\]
in both frameworks.

\subsection{\label{PolosH copy(2)}Other poles of the Lee-Friedrich model.}

Let us now consider the Lee-Friedrich Hamiltonian (\ref{PRE}) for the many
modes sectors, e. g., as an example, for the three mode sector. Then we have
that\footnote{Only symbolically as we have already explained in a previous
footnote.}:%
\begin{equation}
H|z_{1},z_{2},z_{3}\rangle=(z_{1}+z_{2}+z_{3})|z_{1},z_{2},z_{3}%
\rangle\label{cuac-06}%
\end{equation}%
\begin{equation}
H|z_{1},z_{2},z_{0}\rangle=(z_{1}+z_{2}+z_{0})|z_{1},z_{2},z_{0}%
\rangle\label{cuac-07}%
\end{equation}%
\begin{equation}
H|z_{1},z_{0},z_{3}\rangle=(z_{1}+z_{0}+z_{3})|z_{1},z_{0},z_{3}%
\rangle\label{cuac-08}%
\end{equation}%
\begin{equation}
H|z_{0},z_{2},z_{3}\rangle=(z_{0}+z_{2}+z_{3})|z_{0},z_{2},z_{3}%
\rangle\label{cuac-09}%
\end{equation}%
\begin{equation}
H|z_{1},z_{0},z_{0}\rangle=(z_{1}+2z_{0})|z_{1},z_{0},z_{0}\rangle
\label{cuac-10}%
\end{equation}%
\begin{equation}
H|z_{0},z_{2},z_{0}\rangle=\left(  z_{2}+2z_{0}\right)  |z_{0},z_{2}%
,z_{0}\rangle\label{cuac-11}%
\end{equation}%
\begin{equation}
H|z_{0},z_{0},z_{3}\rangle=(z_{3}+2z_{0})|z_{0},z_{0},z_{3}\rangle
\label{cuac-12}%
\end{equation}%
\begin{equation}
H|z_{0},z_{0},z_{0}\rangle=3z_{0}|z_{0},z_{0},z_{0}\rangle\label{cuac-13}%
\end{equation}
where $z_{1},z_{2},z_{3}\epsilon\Gamma.$ So in the real complex plane the
spectrum of $H$ is

1.- From the eigenvalue $(z_{1}+z_{2}+z_{3})$ three points of the curve
$\Gamma$

2.- From the eigenvalue $(z_{1}+z_{2}+z_{0}),$ $(z_{1}+z_{0}+z_{3}),$
$(z_{0}+z_{2}+z_{3})$, a pole at $z_{0\text{ }}$and two points of the curve
$\Gamma$

3.- From the eigenvalue $(z_{1}+2z_{0}),$ $(z_{2}-2z_{0}),$ $(z_{3}+2z_{0})$ a
pole at $2z_{0},$ and one point of the curve $\Gamma$

4.- From the eigenvalue a pole at $3z_{0}$

See figure 3 (\textbf{verificar)}:%

\begin{figure}[t]
\centerline{\scalebox{0.7}{\includegraphics{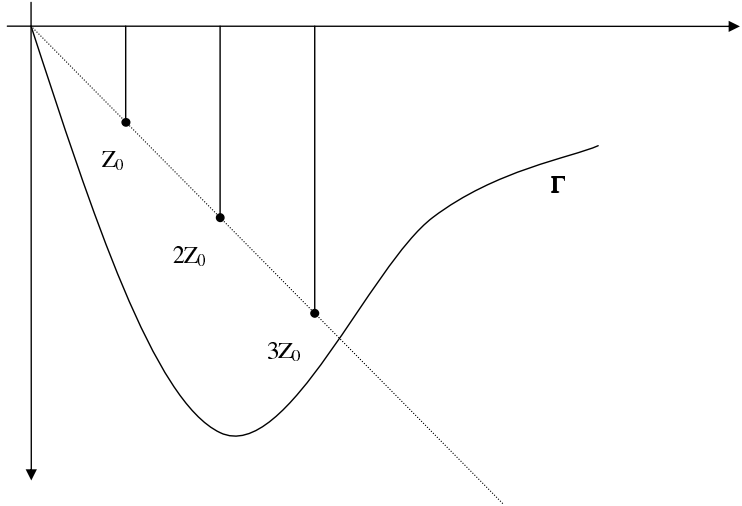}}} \vspace
*{0.cm}\caption{Complex contour on the lower complex energy plane for the
three modes model. The energy poles $z_{0},$ $2z_{0}$ $3z_{0}$ are assumed to
be simple.}%
\label{fig 4}%
\end{figure}

Of course in the general case $3\rightarrow n$ and, as a consequence, the
spectrum is $nz_{0}+$ the curve $\Gamma$, where $n=0,1,2$. $3$,... in fact%
\begin{equation}
z_{n}=nz_{0}+\bigcup_{j}z_{j},\text{ }z_{j}\epsilon\Gamma\label{cuac-14}%
\end{equation}
Then if we neglect the Khalfin term, since it corresponds to extremely long
times, the $\Gamma$ disappears and we simply have%
\begin{equation}
z_{n}=nz_{0}=n(\omega_{0}^{\prime}-i\gamma_{0}) \label{cuac-15}%
\end{equation}
Then under this approximation the system has an effective (non Hermitian)
Hamiltonian%
\[
H_{eff}=\hbar z_{0}a_{0}^{\dagger}a_{0}=\hbar Nz_{0}%
\]
where $a_{0}^{\dagger},a_{0}$ are the creation and annihilation operators for
the mode corresponding to the pole $z_{0}$ and $N$ is the corresponding number
of poles operator. Now the Hamiltonian of the harmonic oscillator is%
\begin{equation}
H_{o}=\left(  N+\frac{1}{2}\right)  \hbar\omega\label{a}%
\end{equation}
Thus we see that in the no Khalfin terms approximation, and taking $\omega
_{0}^{\prime}=\omega$ (of the last equation) and if $n$ is very
large\footnote{Or, in the general case, since $\frac{\hbar\omega}{2}$ only
affects the real part of the pole and not the imaginary namely the one that
produces the time scales.}
\begin{equation}
H_{eff}=H_{o}-i\frac{\gamma_{0}}{\omega}H_{o} \label{b}%
\end{equation}
\ So, in this approximation, the effective Lee-Friedrich Hamiltonian $H_{eff}$
simply is a (non Hermitian) version of $H_{o}$ with a dumping term
$\frac{\gamma_{0}}{\hbar\omega}H_{o}.$ Moreover the basis of $H_{eff}$ and
$H_{o}$ are the same one , i. e.$\{|n\rangle\}$, the basis of the $H_{o}$ of
eq. (\ref{a}).

\subsection{\label{PolosI}Poles dependency on the initial conditions}

\subsubsection{The amplitude of probability}

The probability amplitude that a pure state $\left\vert \varphi\right\rangle $
would be in the pure state $\left\vert \psi\right\rangle $ at time $t$ is:%
\begin{equation}
A(t)=\left\langle \psi|\varphi(t)\right\rangle \label{CI-01}%
\end{equation}
The most general linear superposition of the eigenvectors of $H_{eff}$, in
basis $\{|n\rangle\}$ is:%
\begin{equation}
\left\vert \psi\right\rangle =\sum_{n=0}^{N}a_{n}\left\vert n\right\rangle
\label{CI-02}%
\end{equation}
and the time evolution for $\left\vert \varphi\right\rangle $ must be:%
\begin{equation}
\left\vert \varphi(t)\right\rangle =\sum_{n=0}^{N}b_{n}\left\vert
n(t)\right\rangle \label{CI-03}%
\end{equation}

Then%

\begin{equation}
A(t)=\sum_{n,n^{\prime}=0}^{N}b_{n}a_{n}^{\ast}\left\langle n|n^{\prime
}(t)\right\rangle =\sum_{n,n^{\prime}=0}^{N}b_{n}a_{n^{\prime}}^{\ast
}A_{nn^{\prime}} \label{CI-04}%
\end{equation}
We can compute $A_{nn^{\prime}}=\left\langle n|n^{\prime}(t)\right\rangle
=\left\langle n|\exp\left(  -\frac{i}{\hbar}Ht\right)  |n^{\prime
}\right\rangle =\left\langle n|e^{-iz_{n}t}|n^{\prime}\right\rangle
=e^{-iz_{n}t}\delta_{nn^{\prime}},$ then%

\begin{equation}
A(t)=\sum_{n=0}^{N}b_{n}a_{n}^{\ast}e^{-iz_{n}t} \label{CI-05}%
\end{equation}
where from eqs. (\ref{RT-01}) and (\ref{Omnes}), or eq. 4.47 of
\cite{Manoloetal} we have:
\begin{equation}
z_{n}=\omega_{n}^{\prime}+i\gamma_{n} \label{CI-06}%
\end{equation}
Then, as we know, if we neglect the Khalfin term the "energy" levels are
multiples of the fundamental "energy" i. e.%
\begin{equation}
z_{n}=nz_{0} \label{CI-07}%
\end{equation}
where $z_{0}=\omega_{0}^{\prime}-i\gamma_{0}$ and the coefficients $a_{n}$ and
$b_{n}$\ depend in the initial conditions (according to eq. 4.26 of
\cite{Manoloetal}).

With the expression (\ref{cuac-15}) eq. (\ref{CI-05}) becomes%

\begin{equation}
A(t)=\sum_{n=0}^{N}b_{n}a_{n}^{\ast}e^{-inz_{0}t}=\sum_{n=0}^{N}b_{n}%
a_{n}^{\ast}\left(  e^{-iz_{0}t}\right)  ^{n} \label{CI-09}%
\end{equation}
The same recipe could be used in the fundamental scalar $(\rho_{R}(t)|O_{R})$
instead of $\langle\psi|\varphi(t)\rangle$ with similar results but with more
difficult calculations.

\subsubsection{Initial conditions and evolution}

As initial conditions, $|\alpha_{1}(0)\rangle,$ $|\alpha_{2}(0)\rangle,$ it is
possible to choose any linear combination of the elements $\left\{  \left\vert
n\right\rangle \right\}  $\ with $n=0,1...,\infty$. So we can choose the
coherent states
\begin{equation}
\left\vert \lambda\right\rangle =e^{-\frac{\left\vert \lambda\right\vert ^{2}%
}{2}}\sum_{n=0}^{\infty}\frac{\lambda^{n}}{\sqrt{n!}}\left\vert n\right\rangle
\label{CI-09b}%
\end{equation}
But we can also choose as the \ initial conditions an approximated version
where the number modes is $N$ and we take $n=0,1...,N.$ namely an approximated
quasi-coherent states or quasi-Gaussian (that becomes a coherent state when
$N\rightarrow\infty$ as we will consider below)$.$ Thus
\begin{equation}
\left\vert \lambda\right\rangle =\left(  \sum_{k=0}^{N}\frac{\left\vert
\lambda\right\vert ^{2k}}{k!}\right)  ^{-\frac{1}{2}}\sum_{n=0}^{N}%
\frac{\lambda^{n}}{\sqrt{n!}}\left\vert n\right\rangle \label{CI-09c}%
\end{equation}
Then let us choose the initial conditions as the sum of two quasi-Gaussian
functions, namely:%
\begin{equation}
\left\vert \Phi(0)\right\rangle =a\left\vert \alpha_{1}(0)\right\rangle
+b\left\vert \alpha_{2}(0)\right\rangle \label{CI-10}%
\end{equation}
where $\left\vert \alpha_{1}(0)\right\rangle $ and $\left\vert \alpha
_{2}(0)\right\rangle $ are quasi-coherent states, precisely%
\begin{equation}
\left\vert \alpha_{1}(0)\right\rangle =\left(  \sum_{k=0}^{N}\frac{\left\vert
\alpha_{1}(0)\right\vert ^{2k}}{k!}\right)  ^{-\frac{1}{2}}\sum_{n=0}^{N}%
\frac{\left(  \alpha_{1}(0)\right)  ^{n}}{\sqrt{n!}}\left\vert n\right\rangle
\label{CI-11}%
\end{equation}
and%
\begin{equation}
\left\vert \alpha_{2}(0)\right\rangle =\left(  \sum_{k=0}^{N}\frac{\left\vert
\alpha_{2}(0)\right\vert ^{2k}}{k!}\right)  ^{-\frac{1}{2}}\sum_{n=0}^{N}%
\frac{\left(  \alpha_{2}(0)\right)  ^{n}}{\sqrt{n!}}\left\vert n\right\rangle
\label{CI-12}%
\end{equation}
Thus the initial state is:%
\begin{align}
\rho_{0}  &  =\left\vert \Phi(0)\right\rangle \left\langle \Phi(0)\right\vert
=\left\vert a\right\vert ^{2}\left\vert \alpha_{1}(0)\right\rangle
\left\langle \alpha_{1}(0)\right\vert +ab^{\ast}\left\vert \alpha
_{1}(0)\right\rangle \left\langle \alpha_{2}(0)\right\vert \nonumber\\
&  +a^{\ast}b\left\vert \alpha_{2}(0)\right\rangle \left\langle \alpha
_{1}(0)\right\vert +\left\vert b\right\vert ^{2}\left\vert \alpha
_{2}(0)\right\rangle \left\langle \alpha_{2}(0)\right\vert \label{CI-13}%
\end{align}
Therefore the time evolved state is%
\begin{equation}
\rho(t)=\left\vert \Phi(t)\right\rangle \left\langle \Phi(t)\right\vert
=\rho_{D}(t)+\rho_{ND}(t) \label{CI-14}%
\end{equation}
where $\rho_{D}(t)$\ is the diagonal part (in the basis $\left\{  \left\vert
\alpha_{1}(0)\right\rangle ,\left\vert \alpha_{2}(0)\right\rangle \right\}  $)
of $\rho(t)$
\begin{equation}
\rho^{(D)}(t)=\left\vert a\right\vert ^{2}\left\vert \alpha_{1}%
(t)\right\rangle \left\langle \alpha_{1}(t)\right\vert +\left\vert
b\right\vert ^{2}\left\vert \alpha_{2}(t)\right\rangle \left\langle \alpha
_{2}(t)\right\vert \label{CI-15}%
\end{equation}
and $\rho^{(ND)}$\ is the non-diagonal part of $\rho(t)$%
\begin{equation}
\rho^{(ND)}(t)=ab^{\ast}\left\vert \alpha_{1}(t)\right\rangle \left\langle
\alpha_{2}(t)\right\vert +a^{\ast}b\left\vert \alpha_{2}(t)\right\rangle
\left\langle \alpha_{1}(t)\right\vert \label{CI-16}%
\end{equation}
We choose the two quasi-Gaussian (\ref{CI-11}) and (\ref{CI-12}) with center
at $p_{1,2}(0)=0$, (see \cite{Omnesazul} eq. (7.15) page 284) and%
\begin{equation}
\alpha_{1}(0)=\frac{m\omega}{\sqrt{2m\hbar\omega}}x_{1}(0) \label{CI-17}%
\end{equation}%
\begin{equation}
\alpha_{2}(0)=\frac{m\omega}{\sqrt{2m\hbar\omega}}x_{2}(0) \label{CI-18}%
\end{equation}
So $\alpha_{1}(0)$ and $\alpha_{2}(0)$ are real numbers.

Without loss of generality (since with a change of coordinates we can shift
$x_{1}(0)$ and $x_{2}(0))$ we can consider that the $\alpha_{1}(0)$ and
$\alpha_{2}(0)$ are both positive. For this reason we will interchange
$\alpha_{i}(0)$ and $\left\vert \alpha_{i}(0)\right\vert $ below.

\subsubsection{Components of the non-diagonal part of the state and the
macroscopic case}

Let us not consider $\rho^{(ND)}(t)$ in the basis of the initial condition
$\left\{  \left\vert \alpha_{1}(0)\right\rangle ,\left\vert \alpha
_{2}(0)\right\rangle \right\}  .$Then we have%
\begin{align}
\rho^{(ND)}(t)  &  =\rho_{11}^{(ND)}(t)\left\vert \alpha_{1}(0)\right\rangle
\left\langle \alpha_{1}(0)\right\vert +\rho_{12}^{(ND)}(t)\left\vert
\alpha_{1}(0)\right\rangle \left\langle \alpha_{2}(0)\right\vert \nonumber\\
&  +\rho_{21}^{(ND)}(t)\left\vert \alpha_{2}(0)\right\rangle \left\langle
\alpha_{1}(0)\right\vert +\rho_{22}^{(ND)}(t)\left\vert \alpha_{2}%
(0)\right\rangle \left\langle \alpha_{2}(0)\right\vert \label{CI-20}%
\end{align}

We will prove that for macroscopic initial conditions, i.e. when the peaks of
the two Gaussians are far from each other, the states $\left\{  \left\vert
\alpha_{1}(0)\right\rangle ,\left\vert \alpha_{2}(0)\right\rangle \right\}
$\ are quasi-orthogonal basis, i.e.%
\begin{equation}
\left\langle \alpha_{1}(0)|\alpha_{2}(0)\right\rangle \cong\left\langle
\alpha_{2}(0)|\alpha_{1}(0)\right\rangle \cong0 \label{CI-21}%
\end{equation}
and indeed this is the macroscopicity condition. In fact%
\begin{equation}
\left\langle \alpha_{1}(0)|\alpha_{2}(0)\right\rangle =\left(  \sum_{k=0}%
^{N}\frac{\left\vert \alpha_{1}(0)\right\vert ^{2k}}{k!}\right)  ^{-\frac
{1}{2}}\left(  \sum_{k=0}^{N}\frac{\left\vert \alpha_{2}(0)\right\vert ^{2k}%
}{k!}\right)  ^{-\frac{1}{2}}\sum_{n=0}^{N}\frac{\left(  \alpha_{1}%
(0)\alpha_{2}(0)\right)  ^{n}}{n!} \label{CI-21b}%
\end{equation}
So using the Cauchy product and the binomial theorem we have%
\begin{equation}
\left\langle \alpha_{1}(0)|\alpha_{2}(0)\right\rangle =\left(  \sum_{k=0}%
^{N}\frac{\left(  \left\vert \alpha_{1}(0)\right\vert ^{2}+\left\vert
\alpha_{2}(0)\right\vert ^{2}\right)  ^{k}}{k!}\right)  ^{-\frac{1}{2}}%
\sum_{n=0}^{N}\frac{\left(  \alpha_{1}(0)\alpha_{2}(0)\right)  ^{n}}{n!}
\label{CI-21c}%
\end{equation}
and again using the Cauchy product and the binomial theorem we have%
\begin{equation}
\left\langle \alpha_{1}(0)|\alpha_{2}(0)\right\rangle =\sum_{n=0}^{N}\frac
{1}{n!}\left(  -\frac{1}{2}\left(  \left\vert \alpha_{1}(0)\right\vert
^{2}+\left\vert \alpha_{2}(0)\right\vert ^{2}-2\alpha_{1}(0)\alpha
_{2}(0)\right)  \right)  ^{n} \label{CI-21d}%
\end{equation}
then%
\begin{equation}
\left\langle \alpha_{1}(0)|\alpha_{2}(0)\right\rangle =\sum_{n=0}^{N}\frac
{1}{n!}\left(  -\frac{\left(  \alpha_{1}(0)-\alpha_{2}(0)\right)  ^{2}}%
{2}\right)  ^{n} \label{CI-21e}%
\end{equation}
so for
$\vert$%
$\alpha_{1}(0)-\alpha_{2}(0)|\rightarrow\infty$ we have orthogonality as we
have promised to demonstrate.

In fact, we can consider the limit $N\rightarrow\infty$ . Thus the last scalar
product is equal to the truncated Taylor series of exponential function. Then
we may introduce $R_{N+1},$ the difference with the complete Taylor series,
and we obtain%
\begin{equation}
\left\langle \alpha_{1}(0)|\alpha_{2}(0)\right\rangle =e^{-\frac{\left(
\alpha_{1}(0)-\alpha_{2}(0)\right)  ^{2}}{2}}-R_{N+1} \label{CI-21f}%
\end{equation}
where $R_{N+1}$\ is a correction of order $N+1$
\begin{equation}
R_{N+1}=\frac{e^{\xi}}{\left(  N+1\right)  !}\left(  -\frac{\left(  \alpha
_{1}(0)-\alpha_{2}(0)\right)  ^{2}}{2}\right)  ^{N+1} \label{CI-21g}%
\end{equation}
with $\xi\in\left(  -\frac{\left(  \alpha_{1}(0)-\alpha_{2}(0)\right)  ^{2}%
}{2},0\right)  $, then we have%
\begin{equation}
R_{N+1}\leq\frac{1}{\left(  N+1\right)  !}\left(  -\frac{\left(  \alpha
_{1}(0)-\alpha_{2}(0)\right)  ^{2}}{2}\right)  ^{N+1} \label{CI-21h}%
\end{equation}
Thus we can prove the orthogonality conditions:

\begin{enumerate}
\item To eliminate the first term of (\ref{CI-21f}) we make%
\begin{equation}
e^{-\frac{\left(  \alpha_{1}(0)-\alpha_{2}(0)\right)  ^{2}}{2}}\ll1
\label{CI-21i}%
\end{equation}
i.e.
\begin{equation}
\left\vert \alpha_{1}(0)-\alpha_{2}(0)\right\vert \gg1 \label{CI-21j}%
\end{equation}

\item To eliminate the second term of (\ref{CI-21f}), $\left\vert
R_{N+1}\right\vert \ll1$, we make%
\begin{equation}
\left\vert \frac{1}{\left(  N+1\right)  !}\left(  -\frac{\left(  \alpha
_{1}(0)-\alpha_{2}(0)\right)  ^{2}}{2}\right)  ^{N+1}\right\vert \ll1
\label{CI-21k}%
\end{equation}
i.e.%
\begin{equation}
\left\vert \alpha_{1}(0)-\alpha_{2}(0)\right\vert \ll\left[  2\left(
N+1\right)  !\right]  ^{\frac{1}{2\left(  N+1\right)  }} \tag{CI-21l}%
\end{equation}

This expression can be simplified by a huge $N$ using the Stirling's
approximation%
\begin{equation}
\left\vert \alpha_{1}(0)-\alpha_{2}(0)\right\vert \ll\sqrt{2\left(
N+1\right)  } \label{CI-21m}%
\end{equation}
Then the orthogonality is proved if eqs. (\ref{CI-21j}) and (\ref{CI-21m}) are satisfy.

In fact these are the two \textit{macroscopicity} condition: then $\left\vert
\alpha_{1}(0)-\alpha_{2}(0)\right\vert $ and $N$ should be large.
\end{enumerate}

We will consider that $\alpha_{1}(0)-\alpha_{2}(0)$ and $N+1$ always satisfy
these macroscopicity conditions. Then from (\ref{CI-21f}) the basis $\left\{
\left\vert \alpha_{1}(0)\right\rangle ,\left\vert \alpha_{2}(0)\right\rangle
\right\}  $ is quasi-orthogonal and we have
\begin{align}
\rho_{11}^{(ND)}(t)  &  =\left\langle \alpha_{1}(0)\right\vert \rho
^{(ND)}(t)\left\vert \alpha_{1}(0)\right\rangle \nonumber\\
\rho_{12}^{(ND)}(t)  &  =\left\langle \alpha_{1}(0)\right\vert \rho
^{(ND)}(t)\left\vert \alpha_{2}(0)\right\rangle \nonumber\\
\rho_{21}^{(ND)}(t)  &  =\left\langle \alpha_{2}(0)\right\vert \rho
^{(ND)}(t)\left\vert \alpha_{1}(0)\right\rangle \nonumber\\
\rho_{22}^{(ND)}(t)  &  =\left\langle \alpha_{2}(0)\right\vert \rho
^{(ND)}(t)\left\vert \alpha_{2}(0)\right\rangle \label{CI-22}%
\end{align}
then from eq. (\ref{CI-16}) we have%
\begin{align}
\rho_{11}^{(ND)}(t)  &  =ab^{\ast}\left\langle \alpha_{1}(0)|\alpha
_{1}(t)\right\rangle \left\langle \alpha_{2}(t)|\alpha_{1}(0)\right\rangle
+a^{\ast}b\left\langle \alpha_{1}(0)|\alpha_{2}(t)\right\rangle \left\langle
\alpha_{1}(t)|\alpha_{1}(0)\right\rangle \nonumber\\
\rho_{12}^{(ND)}(t)  &  =ab^{\ast}\left\langle \alpha_{1}(0)|\alpha
_{1}(t)\right\rangle \left\langle \alpha_{2}(t)|\alpha_{2}(0)\right\rangle
+a^{\ast}b\left\langle \alpha_{1}(0)|\alpha_{2}(t)\right\rangle \left\langle
\alpha_{1}(t)|\alpha_{2}(0)\right\rangle \nonumber\\
\rho_{21}^{(ND)}(t)  &  =ab^{\ast}\left\langle \alpha_{2}(0)|\alpha
_{1}(t)\right\rangle \left\langle \alpha_{2}(t)|\alpha_{1}(0)\right\rangle
+a^{\ast}b\left\langle \alpha_{2}(0)|\alpha_{2}(t)\right\rangle \left\langle
\alpha_{1}(t)|\alpha_{1}(0)\right\rangle \nonumber\\
\rho_{22}^{(ND)}(t)  &  =ab^{\ast}\left\langle \alpha_{2}(0)|\alpha
_{1}(t)\right\rangle \left\langle \alpha_{2}(t)|\alpha_{2}(0)\right\rangle
+a^{\ast}b\left\langle \alpha_{2}(0)|\alpha_{2}(t)\right\rangle \left\langle
\alpha_{1}(t)|\alpha_{2}(0)\right\rangle \label{CI-23}%
\end{align}
We can compute these products with eqs. (\ref{cuac-15}) and (\ref{CI-09}).

\begin{description}
\item[-] For $\left\langle \alpha_{1}(0)|\alpha_{1}(t)\right\rangle $ we have
that $\left\vert \psi\right\rangle =\left\vert \alpha_{1}(0)\right\rangle $
and $\left\vert \varphi(t)\right\rangle =\left\vert \alpha_{1}(t)\right\rangle
$, then from (\ref{CI-11}) and since $\alpha_{1}(t)$ is a real number
\[
a_{n}^{\ast}=e^{-\frac{\left\vert \alpha_{1}(0)\right\vert ^{2}}{2}}%
\frac{\left(  \alpha_{1}(0)\right)  ^{n}}{\sqrt{n!}}andb_{n}=e^{-\frac
{\left\vert \alpha_{1}(0)\right\vert ^{2}}{2}}\frac{\left(  \alpha
_{1}(0)\right)  ^{n}}{\sqrt{n!}}%
\]

\item then we have%
\begin{align}
\left\langle \alpha_{1}(0)|\alpha_{1}(t)\right\rangle  &  =e^{-\left\vert
\alpha_{1}(0)\right\vert ^{2}}\sum_{n=0}^{N}\frac{\left(  \left\vert
\alpha_{1}(0)\right\vert ^{2}\right)  ^{n}}{n!}\left(  e^{-iz_{0}t}\right)
^{n}\nonumber\\
&  =e^{-\left\vert \alpha_{1}(0)\right\vert ^{2}}e^{\left\vert \alpha
_{1}(0)\right\vert ^{2}e^{-iz_{0}t}} \label{CI-25}%
\end{align}

\item[-] For $\left\langle \alpha_{1}(0)|\alpha_{2}(t)\right\rangle $ we have
that $\left\vert \psi\right\rangle =\left\vert \alpha_{1}(0)\right\rangle $
and $\left\vert \varphi(t)\right\rangle =\left\vert \alpha_{2}(t)\right\rangle
$, then from (\ref{CI-11}), (\ref{CI-12}) and since $\alpha_{1}(t)$ and
$\alpha_{2}(t)$ are real numbers we can find the coefficients of eqs.
(\ref{CI-02}) and (\ref{CI-03}) as
\begin{equation}
a_{n}^{\ast}=e^{-\frac{\left\vert \alpha_{1}(0)\right\vert ^{2}}{2}}%
\frac{\left\vert \alpha_{1}(0)\right\vert ^{n}}{\sqrt{n!}}\text{ \ \ and
\ }b_{n}=e^{-\frac{\left\vert \alpha_{2}(0)\right\vert ^{2}}{2}}%
\frac{\left\vert \alpha_{2}(0)\right\vert ^{n}}{\sqrt{n!}} \label{CI-26}%
\end{equation}
then, form eq. (\ref{CI-05}) we have%
\begin{align}
\left\langle \alpha_{1}(0)|\alpha_{2}(t)\right\rangle  &  =e^{-\frac
{\left\vert \alpha_{1}(0)\right\vert ^{2}+\left\vert \alpha_{2}(0)\right\vert
^{2}}{2}}\sum_{n=0}^{N}\frac{\left(  \left\vert \alpha_{1}(0)\right\vert
\left\vert \alpha_{2}(0)\right\vert \right)  ^{n}}{n!}\left(  e^{-iz_{0}%
t}\right)  ^{n}\nonumber\\
&  =e^{-\frac{\left\vert \alpha_{1}(0)\right\vert ^{2}+\left\vert \alpha
_{2}(0)\right\vert ^{2}}{2}}e^{\left\vert \alpha_{1}(0)\right\vert \left\vert
\alpha_{2}(0)\right\vert e^{-iz_{0}t}} \label{CI-27}%
\end{align}

\item[-] For $\left\langle \alpha_{2}(0)|\alpha_{1}(t)\right\rangle $ we have
that $\left\vert \psi\right\rangle =\left\vert \alpha_{2}(0)\right\rangle $
and $\left\vert \varphi(t)\right\rangle =\left\vert \alpha_{1}(t)\right\rangle
$, then from (\ref{CI-11}), (\ref{CI-12}) and since $\alpha_{1}(t)$ and
$\alpha_{2}(t)$ are real numbers
\begin{equation}
a_{n}^{\ast}=e^{-\frac{\left\vert \alpha_{2}(0)\right\vert ^{2}}{2}}%
\frac{\left(  \alpha_{2}(0)\right)  ^{n}}{\sqrt{n!}}\text{ and }%
b_{n}=e^{-\frac{\left\vert \alpha_{1}(0)\right\vert ^{2}}{2}}\frac{\left(
\alpha_{1}(0)\right)  ^{n}}{\sqrt{n!}} \label{CI-28}%
\end{equation}
then%
\begin{align}
\left\langle \alpha_{2}(0)|\alpha_{1}(t)\right\rangle  &  =e^{-\frac
{\left\vert \alpha_{1}(0)\right\vert ^{2}+\left\vert \alpha_{2}(0)\right\vert
^{2}}{2}}\sum_{n=0}^{N}\frac{\left(  \left\vert \alpha_{1}(0)\right\vert
\left\vert \alpha_{2}(0)\right\vert \right)  ^{n}}{n!}\left(  e^{-iz_{0}%
t}\right)  ^{n}\nonumber\\
&  =e^{-\frac{\left\vert \alpha_{1}(0)\right\vert ^{2}+\left\vert \alpha
_{2}(0)\right\vert ^{2}}{2}}e^{\left\vert \alpha_{1}(0)\right\vert \left\vert
\alpha_{2}(0)\right\vert e^{-iz_{0}t}} \label{CI-29}%
\end{align}

\item[-] For $\left\langle \alpha_{2}(0)|\alpha_{2}(t)\right\rangle $ we have
that $\left\vert \psi\right\rangle =\left\vert \alpha_{2}(0)\right\rangle $
and $\left\vert \varphi(t)\right\rangle =\left\vert \alpha_{2}(t)\right\rangle
$, then from (\ref{CI-12}) and since $\alpha_{2}(t)$ is a real number
\begin{equation}
a_{n}^{\ast}=e^{-\frac{\left\vert \alpha_{2}(0)\right\vert ^{2}}{2}}%
\frac{\left(  \alpha_{2}(0)\right)  ^{n}}{\sqrt{n!}}\text{ and }%
b_{n}=e^{-\frac{\left\vert \alpha_{2}(0)\right\vert ^{2}}{2}}\frac{\left(
\alpha_{2}(0)\right)  ^{n}}{\sqrt{n!}} \label{CI-30}%
\end{equation}
then%
\begin{align}
\left\langle \alpha_{2}(0)|\alpha_{2}(t)\right\rangle  &  =e^{-\left\vert
\alpha_{2}(0)\right\vert ^{2}}\sum_{n=0}^{N}\frac{\left(  \left\vert
\alpha_{2}(0)\right\vert ^{2}\right)  ^{n}}{n!}\left(  e^{-iz_{0}t}\right)
^{n}\nonumber\\
&  =e^{-\left\vert \alpha_{2}(0)\right\vert ^{2}}e^{\left\vert \alpha
_{2}(0)\right\vert ^{2}e^{-iz_{0}t}} \label{CI-31}%
\end{align}

\end{description}

Now if we consider eqs. (\ref{CI-17}) and (\ref{CI-18}) and remember that the
initial centers of the Gaussians are given by eqs. (\ref{CI-11}) and
(\ref{CI-12}), with no lost of generality we can choose:
\begin{equation}
\alpha_{1}(0)=0 \label{CI-32}%
\end{equation}
and%
\begin{equation}
\alpha_{2}(0)=\frac{m\omega}{\sqrt{2m\hbar\omega}}L_{0} \label{CI-33}%
\end{equation}

Remember that we have imposed a macroscopic condition to the initial
conditions, i.e. $\left\vert \alpha_{1}(0)-\alpha_{2}(0)\right\vert \gg1$ and
$\left\vert \alpha_{1}(0)-\alpha_{2}(0)\right\vert \ll\left[  2\left(
N+1\right)  !\right]  ^{\frac{1}{2\left(  N+1\right)  }}$. So in the case
given by (\ref{CI-32}) and (\ref{CI-33}) we have%
\begin{equation}
\left\vert \alpha_{1}(0)-\alpha_{2}(0)\right\vert =\alpha_{2}(0)\gg1\text{
\ \ and \ \ }\left\vert \alpha_{1}(0)-\alpha_{2}(0)\right\vert \ll\left[
2\left(  N+1\right)  !\right]  ^{\frac{1}{2\left(  N+1\right)  }}
\label{CI-34}%
\end{equation}
i.e.%
\begin{equation}
\frac{m\omega}{\sqrt{2m\hbar\omega}}L_{0}\gg1\text{ \ \ and \ }\left[
2\left(  N+1\right)  !\right]  ^{\frac{1}{2\left(  N+1\right)  }}\gg
\text{\ }\frac{m\omega}{\sqrt{2m\hbar\omega}}L_{0} \label{CI-34b}%
\end{equation}
Then if we substitute (\ref{CI-32}), (\ref{CI-33}) and (\ref{CI-34}) in eq.
(\ref{CI-25}), (\ref{CI-27}), (\ref{CI-29}) and (\ref{CI-31}) and we take into
account (\ref{CI-34})%
\begin{equation}
\left\langle \alpha_{1}(0)|\alpha_{1}(t)\right\rangle =1 \label{CI-35}%
\end{equation}%
\begin{equation}
\left\langle \alpha_{1}(0)|\alpha_{2}(t)\right\rangle =e^{-\frac{\left\vert
\alpha_{2}(0)\right\vert ^{2}}{2}}\cong0 \label{CI-36}%
\end{equation}%
\begin{equation}
\left\langle \alpha_{2}(0)|\alpha_{1}(t)\right\rangle =e^{-\frac{\left\vert
\alpha_{2}(0)\right\vert ^{2}}{2}}\cong0 \label{CI-37}%
\end{equation}%
\begin{equation}
\left\langle \alpha_{2}(0)|\alpha_{2}(t)\right\rangle =e^{-\left\vert
\alpha_{2}(0)\right\vert ^{2}\left(  1-e^{-iz_{0}t}\right)  } \label{CI-38}%
\end{equation}
Moreover if we substitute (\ref{CI-35}), (\ref{CI-36}), (\ref{CI-37}) and
(\ref{CI-38}) in eq. (\ref{CI-23}) we have%
\begin{align}
\rho_{11}^{(ND)}(t)  &  \cong0\nonumber\\
\rho_{12}^{(ND)}(t)  &  \cong ab^{\ast}e^{-\left\vert \alpha_{2}(0)\right\vert
^{2}\left(  1-e^{-iz_{0}t}\right)  }\nonumber\\
\rho_{21}^{(ND)}(t)  &  \cong a^{\ast}be^{-\left\vert \alpha_{2}(0)\right\vert
^{2}\left(  1-e^{-iz_{0}t}\right)  }\nonumber\\
\rho_{22}^{(ND)}(t)  &  \cong0 \label{CI-39}%
\end{align}

We see that in these equations there is an exponential produced by the poles
of eq. (\ref{CI-07}), Then we have from eq. (\ref{CI-20}) and eq.
(\ref{CI-33}).%
\begin{equation}
\rho_{ij}^{(ND)}(t)\propto\exp[-\frac{m\omega}{2\hbar}L_{0}^{2}(1-e^{-\gamma
_{0}t}) \label{CI-40}%
\end{equation}
Then we can make the expansion we have done in eq. (\ref{app}).%
\[
\rho_{ij}^{(ND)}(t)\propto\exp[-\frac{m\omega}{2\hbar}L_{0}^{2}(1-1+\gamma
_{0}t-\frac{1}{2}\gamma_{0}^{2}t^{2}-...)]
\]
and introduce the recipe of eq. (\ref{tesis}) to obtain
\[
\rho_{ij}^{(ND)}(t)\propto\exp(-\frac{m\omega}{2\hbar}L_{0}^{2}\gamma_{0}t)
\]
So a largest decaying time $t_{R}=\frac{1}{\gamma_{0}}$ is given by the
original pole of eq. (\ref{cuac-03}) but, using our poles technique, a new
decaying mode appears, as in eq. (\ref{Omneslike}), i.e.%
\[
f(t)\sim\exp(-\frac{\gamma_{eff}}{\hbar}t)
\]
with the coefficient $\gamma_{eff}$ that, with the convention of this section,
reads:
\begin{equation}
\gamma_{eff}=\frac{m\omega}{2\hbar}L_{0}^{2}\gamma_{0} \label{CI-41}%
\end{equation}
so, the new characteristic time is $t_{D}=\frac{1}{\gamma_{eff}}$ or
\begin{equation}
t_{D}=\frac{2\hbar}{m\omega}\frac{1}{L_{0}^{2}}t_{R} \label{CI-42}%
\end{equation}

the same time was found by Omn\`{e}s in \cite{Omnesazul} or in eq.
(\ref{131'}) and corresponding to the definition (\ref{34'}) of section III.A.
So in fact, we have recovered the same result. Also in \cite{Omnesazul} the
result for $t_{D}$ is valid only for small $t$ as in section III.A. In the
general case, and considering that $\alpha_{1}(0)=0$, from eqs. (\ref{CI-20})
and (\ref{CI-40})\ we have:%
\begin{align}
\rho^{(ND)}(t)  &  =\left\{  ab^{\ast}\left\vert \alpha_{1}(0)\right\rangle
\left\langle \alpha_{2}(0)\right\vert +ba^{\ast}\left\vert \alpha
_{1}(0)\right\rangle \left\langle \alpha_{2}(0)\right\vert \right\}
\nonumber\\
&  \exp\left[  -\frac{m\omega}{2\hbar}L_{0}^{2}\left(  1-e^{-\gamma_{0}%
t}\right)  \right]  \label{CI-43}%
\end{align}
the same expression that can be found on page 290 of~\cite{Omnesazul} or in
eq. (\ref{131}). So the coincidence of both formalisms is completely proved.

Finally, for $t<t_{D}$ the evolution of $\rho_{R}(t)$ contains all the modes
(more or less as in figure of 3.315$_{a}$ of \cite{Ex}). For times $t>t_{D}$
the fast modes are not important anymore and the evolution corresponds to eq.
(\ref{EV}) where only the slow modes have influence in the evolution of
$\rho_{R}(t)$ (more or less as in figure of 3.315$_{b}$ of \cite{Ex}). Then in
order to define the moving preferred basis we must consider the definition of
$\rho_{R}(t)$ in eq. (\ref{EV}). For larger times only the "big mountains"
\ of figure 3.315$_{b}$ of \cite{Ex} remain while the central interference
pattern has vanished. these mountains in motion are the moving preferred basis
in this case. Finally for $t\sim t_{R}$ the two mountains have reached
equilibrium and only the $\rho_{D}(t)$ remains. This would we the evolution of
the moving preferred basis of Omn\`{e}s (even if he never use this name)
explained according to our formalism.

So we have proved that all the characters of the Omn\`{e}s model: $t_{R},$
$t_{D}$, and the moving preferred basis, coincide with our definitions of
\ section III.A.

Precisely, as we have said in the macroscopic case the basis \{%
$\vert$%
$\alpha_{1}(t)\rangle,|\alpha_{2}(t)\rangle\}$ is orthogonal and it is the one
defined in section III.A. In fact for $t>t_{D}$ the evolution of $\rho_{R}(t)$
is produced by the poles such that $\gamma_{i}<\gamma_{eff}$ while for
$t<t_{D}$ the evolution is produced by all the poles.. Moreover the
corresponding $\rho_{R}(t)$ and $\rho_{P}(t)$ coincide at $t=t_{D},$ with all
their derivatives.

\section{\label{Conclusions}Conclusions}

In this paper we have:

i.- Discussed a general scheme for decoherence, that in principle could be
used in many examples.

ii.- We have given a quite general definition of moving preferred basis
$\{\widetilde{|j(t)\rangle}\}$, and of relaxation and decoherence times for a
general systems.

iii.- We have proved that our definitions coincide with those of the Omn\`{e}s model..

We hope that these general results will produce some light in the general
problem of decoherence.

The Omn\`{e}s formalism, of references \cite{OmnesPh}, \cite{OmnesRojo}, and
\cite{Omnesazul} contains the most general definition of moving preferred
basis of the literature on the subject. Our basis have another conceptual
frame: the catalogue of decaying modes in the non-unitary evolution of a
quantum system. But since the Omn\`{e}s formalism is the best available it
would be very important for us to show, in the future, the coincidence of both
formalisms, as we have done at least for one model in this paper (see section III.B).

Of course we realize that, to prove our proposal, more examples must be added,
as we will do elsewhere. But we also believe that we have a good point of
depart. In fact, probably the coincidences that we have found in the Omn\`{e}s
model could be a general feature of the decoherence phenomenon. Essentially
because, being the poles catalogue the one that contains \textit{all the
possible decaying modes} of the non unitary evolutions, since relaxation and
decoherence are non-unitary evolutions, necessarily they must be contained
within this catalogue, .

\section{Acknowledgments}

We are very grateful to Roberto Laura, Olimpia Lombardi, Roland Omn\`{e}s and
Maximilian Schlosshauer for many comments and criticisms. This research was
partially supported by grants of the University of Buenos Aires, the CONICET
and the FONCYT of Argentina.

\section{Appendix A.}

\subsection{Observables that see some poles.}

In this appendix we will introduce a particular example of observables, of the
same system, such that some observables would see some poles while other would
see other poles. Essentially it is a bi-Friedrich-model.

Let us consider a system $\mathcal{S}$ with Hamiltonian:%
\[
H=H_{0}+H_{Int}%
\]
where%
\[
H_{0}=\Omega_{1}|1\rangle\langle1|+\Omega_{2}|2\rangle\langle2|+2\int
_{0}^{\infty}\omega|\omega\rangle\langle\omega|d\omega
\]
and
\[
H_{Int}=\int_{0}^{a}V_{\omega}^{(1)}\left[  |\omega\rangle\langle
1|+|1\rangle\langle\omega|\right]  d\omega+\int_{b}^{\infty}V_{\omega^{\prime
}}^{(2)}\left[  |\omega^{\prime}\rangle\langle2|+|2\rangle\langle
\omega^{\prime}|\right]  d\omega
\]
where $a<b$ and $\langle1|2\rangle=\langle\omega|2\rangle=\langle
1|\omega\rangle=0.$ This Hamiltonian can also reads:%
\[
H=H_{1}+H_{2}%
\]
where%
\[
H_{1}=\Omega_{1}|1\rangle\langle1|+\int_{0}^{\infty}\omega|\omega
\rangle\langle\omega|d\omega+\int_{0}^{a}V_{\omega}^{(1)}\left[
|\omega\rangle\langle2|+|2\rangle\langle\omega|\right]  d\omega
\]
and%
\[
H_{2}=\Omega_{2}|2\rangle\langle2|+\int_{0}^{\infty}\omega^{\prime}%
|\omega^{\prime}\rangle\langle\omega^{\prime}|d\omega^{\prime}+\int
_{b}^{\infty}V_{\omega^{\prime}}^{(2)}\left[  |\omega^{\prime}\rangle
\langle1|+|1\rangle\langle\omega^{\prime}|\right]  d\omega^{\prime}%
\]
$.$ Then it is easy to prove that%
\[
\lbrack H_{1},H_{2}]=0
\]
and that%
\[
\exp(-\frac{i}{\hbar}Ht)=\exp(-\frac{i}{\hbar}H_{1}t)\exp(-\frac{i}{\hbar
}H_{2}t)
\]
Let us now decompose the system as $\mathcal{S=P}_{1}\cup\mathcal{P}_{2}$
\ where part $\mathcal{P}_{1}$ is related with Hamiltonian $H_{1}$ and part
$\mathcal{P}_{2}$ related with Hamiltonian $H_{2}$. Let us observe that these
two parts are not independent since they share a common continuous spectrum,
i. e. $2\int_{0}^{\infty}\omega|\omega\rangle\langle\omega|d\omega$. Moreover
let the corresponding relevant observable spaces be $\mathcal{O}_{1}\otimes
I_{E1}$ for $\mathcal{P}_{1}$ and $\mathcal{O}_{2}\otimes I_{E2}$ for
$\mathcal{P}_{2},$ where $\mathcal{O}_{1}$ has basis \{$|1\rangle\},$ and
$\mathcal{O}_{E1}$ has basis \{$|\omega\rangle\}$ while $\mathcal{O}_{2}$ has
basis \{$|2\rangle\},$ and $\mathcal{O}_{E2}$ basis \{$|\omega^{\prime}%
\rangle\}.$ Moreover let us consider the two relevant observables of system
$\mathcal{S=P}_{1}\cup\mathcal{P}_{2}$
\[
\mathbb{O}_{i}=O_{1}\otimes I_{E1}\otimes\mathcal{I}_{2}\otimes I_{E2}\text{
and }\mathbb{O}_{2}=I_{1}\otimes I_{E1}\otimes O_{2}\otimes I_{E2}%
\]
where the $I$ are the corresponding unit operators. Then
\[
(\rho(t)|\mathbb{O}_{1})=(\rho(0)|\exp(\frac{i}{\hbar}H_{1}t)O_{1}\exp
(-\frac{i}{\hbar}H_{1}t)\otimes I_{E1}\otimes\mathcal{I}_{2}\otimes I_{E2})
\]%
\[
(\rho(t)|\mathbb{O}_{2})=(\rho(0)|I_{1}\otimes I_{E1}\otimes\exp(\frac
{i}{\hbar}H_{2}t)O_{2}\exp(\frac{i}{\hbar}H_{2}t)\otimes I_{E2})
\]
and therefore $\mathbb{O}_{1}$ only sees the evolution in part $\mathcal{P}%
_{1}$ while $\mathbb{O}_{2}$ only sees the evolution in part $\mathcal{P}%
_{2}.$ Then, since the poles of part $\mathcal{P}_{1}$ correspond to the
decaying modes of the evolution of this part (and we know that the Friedrich
model of this subsystem generically do have poles) $\mathbb{O}_{1}$ only sees
the poles of part $\mathcal{P}_{1}$. Respectively $\mathbb{O}_{2}$ only sees
the poles of part $\mathcal{P}_{2}.$ q. e. d.

Now we can consider that the poles of part $\mathcal{P}_{1}$ define a
relaxation time $t_{R1\text{ }}$while the poles of part $\mathcal{P}_{2}$
define a relaxation time $t_{R2\text{ }}.$ If $t_{R1\text{ }}\ll t_{R2\text{
}}$part $\mathcal{P}_{1}$ decoheres and becomes classical in a\ short time
$t>t_{R1\text{ }}$while part $\mathcal{P}_{2}$ remains quantum for a large
time $t_{\text{ }}<t_{R2\text{ }}.$ Then for $t$ such that $t_{R1\text{ }%
}<t<t_{R2\text{ }}$ part $\mathcal{P}_{1}$ behaves classically while part
$\mathcal{P}_{2}$ remains quantum. Precisely: system $\mathcal{S}$ observed by
$\mathbb{O}_{i}=O_{1}\otimes I_{E1}\otimes\mathcal{I}_{2}\otimes I_{E2}$ seems
classical while observed by $\mathbb{O}_{2}=I_{1}\otimes I_{E1}\otimes
O_{2}\otimes I_{E2}$ \ seems quantum. In fact this is the behavior of a
generic physical system.

\end{document}